\documentclass[%
 reprint,
 amsmath,amssymb,
 aps,
]{revtex4-2}

\usepackage{graphicx}
\usepackage{dcolumn}
\usepackage{bm}
\usepackage{newtxtext,newtxmath}
\usepackage{natbib}

\begin{document}

\preprint{APS/123-QED}

\title{Kiloparsec-scale turbulence driven by reionization may grow intergalactic magnetic fields}

\author{Christopher Cain}
\email{clcain3@asu.edu}
\affiliation{School of Earth and Space exploration, Arizona State University, Tempe, AZ 85281, USA}

\author{Matthew McQuinn}
 \email{mcquinn@uw.edu}
\affiliation{Department of Astronomy, University of Washington, Seattle, WA 98195-1580, USA}

\author{Evan Scannapieco}
\email{evan.scannapieco@asu.edu }
\affiliation{School of Earth and Space exploration, Arizona State University, Tempe, AZ 85281, USA }

\author{Anson D'Aloisio}
\email{ansond@ucr.edu}
\affiliation{Department of Physics and Astronomy, University of California, Riverside, CA 92521, USA}

\author{Hy Trac}
\email{hytrac@andrew.cmu.edu }
\affiliation{McWilliams Center for Cosmology and Astrophysics, Department of Physics, Carnegie Mellon University, Pittsburgh, PA 15213, USA}

\date{\today}

\begin{abstract}

The intergalactic medium (IGM) underwent intense heating that resulted in pressure disequilibrium in the wake of ionization fronts during cosmic reionization.  The dynamical relaxation to restore pressure balance may have driven small-scale turbulence and, hence, the amplification of intergalactic magnetic fields.  We investigate this possibility for the first time using a suite of $\approx 100$ pc resolution radiation-hydrodynamics simulations of IGM gas dynamics.  We show that as the spatial resolution improves beyond that achieved with most prior studies, much of the IGM becomes turbulent unless it was pre-heated to $\gg 100~$K before reionization.  
In our most turbulent simulations, we find that the gas energy spectrum follows the expected $k^{-5/3}$ Kolmogorov scaling to the simulation's resolution, and the eddy turnover time of the turbulence is $< 1$ Gyr at $k \approx 1 ~$kpc$^{-1}$.  Turbulence will grow magnetic fields, and we show that the fields grown by reionization-driven turbulence could explain lower limits on the strength of volume-filling B-fields from observations of TeV blazars.  As reionization sweeps over the cosmos, this mechanism could create turbulence throughout the cosmic volume with a character that only depends on the amount of IGM preheating.  
\end{abstract}

\maketitle


\section{Introduction}

Hydrogen in the Intergalactic Medium (IGM) underwent a major phase transition about a billion years after the big bang, when it was ionized by the first generation of galaxies.  This transition not only changed its ionization state, but also its temperature~\citep{McQuinn2016,DAloisio2019,Zeng2021} and small-scale dynamics~\citep{Shapiro2004,Park2016,DAloisio2020,Nasir2021,Chan2023}.  Before reionization, models predict that the IGM temperature was in the range $10$-$1000~$K~\citep{Fialkov2014c,Ross2017,Eide2018}.  For the coldest of these temperatures, the gas may have formed structures as light as $10^4~M_{\odot}$~\citep{Gnedin2000,Emberson2013}.  Reionization impulsively heated this gas to a few $\times 10^4$K, over-pressurizing these systems and causing them to dynamically relax to a scale of tens of kiloparsecs over a few hundred million years \citep{DAloisio2020}. 

Several studies have used high-resolution cosmological simulations to study this relaxation process and how it affects various observables ~\citep{Park2016,Hirata2018,DAloisio2020,Doughty2023,Puchwein2023, Cain2024a, Davies2024}.  Here we investigate whether relaxation could drive turbulence that was unresolved in these prior studies.  There is reason to suspect that the post-reionization IGM may have been turbulent.  After ionization fronts (boundaries between fully neutral and highly ionized gas, I-fronts) sweep through a region, the gas would have been evacuated out of mini-halos and filaments and transported $\sim 10$s of co-moving kiloparsecs~\citep{DAloisio2020,Chan2023}, providing a natural driving mechanism.  As the recently ionized gas expands out of the potential well, it cools adiabatically, reaching temperatures as low as a few thousand K~\citep{Hirata2018,Cain2024a}.  Nearby diffuse gas is compressed by this process, and can reach $5\times10^4~$K or higher within $\approx 100$ Myr of reionization.  The pressure differences between hot and cool gas can then drive turbulence.  This potential mechanism has the unique features that it would be expected to occur everywhere in the IGM (since reionization happens everywhere) and be independent of process at play within galaxies and the circumgalactic medium.  To our knowledge, a mechanism with these properties has never been studied before.  


Turbulent conditions in the IGM during and after reionization may grow intergalactic magnetic fields.  There has been significant previous research on the seeding of IGM magnetic fields.  One of the most referenced seeding mechanisms is the ``Biermann battery'', in which inhomogeneous heating leads to small charge separation and the resulting electric fields generate B fields via Faraday's law.  This can occur during reionization~\citep{Biermann1950,Subramanian1994,Kulsrud2008} or in the pre-recombination epoch~\citep{Quashnock1988,Sigl1997}.  Additionally, there may be seeds generated prior to reionization, such as from the streaming of cosmic rays produced by the first supernovae~\citep{Miniati2011} and possibly even primordially during the Big Bang~\citep{Subramanian2016}.  
However, regardless of whether these seed fields are produced before recombination (``primordial'' seed fields) or afterwards (``astrophysical'' seed fields), their expected amplitudes are $\lesssim 10^{-20}$ G~\citep[e.g.][]{Turner1988,Sigl1997,Durrer2013}, their amplitudes are expected to be small, $\lesssim 10^{-20}$ G.  
Reaching larger field strengths requires an amplification mechanism.  
One such mechanism is the so-called turbulent dynamo, which has been explored in detail both in general setups \citep{Schekochihin2004,Xu2016, Rincon_2019} and, e.g. in the context of forming proto-galaxies~\citep{Kulsrud1997, Balsara_2004, Xu2016}.  Here we consider the turbulent dynamo in the context of reionization-driven turbulence.   

Explaining the growth of volume-filling IGM magnetic fields is of particular interest in light of lower limits on the strength of these fields from TeV blazars~\citep{Neronov2010,AlvesBatista2021}.  In the absence of magnetic fields, one expects to observe $\gamma$-ray halos around these objects generated by the inverse-Compton scattering of TeV photons off of CMB photons.  The absence of these halos both observed in the spectral energy distribution and the spatial extent of the sources has been interpreted as an effect of deflection by IGM magnetic fields~(an alternative plasma instability hypothesis has been recently ruled out \footnote{An alternative explanation of the blazar observations is that the TeV pairs lose their energy owing to plasma beam instabilities \citep{Broderick2012}. After an extensive debate, this possibility appears to have been ruled out. It was recently shown that Landau damping by non-thermal electrons generated by Compton scattering off of the gamma ray background should suppress the growth of these instabilities \citep{Yang2024}.}).  This absence places a conservative lower limit on their field strength of $B > 3 \times 10^{-16}$ G for fields correlated over $10$ kpc~\citep{2010Sci...328...73N, 2011A&A...529A.144T, 2018ApJS..237...32A}.  This explanation requires significant growth of seed fields, whatever their origin.  It also requires that the mechanism responsible for driving turbulence be effective everywhere in the IGM to explain Blazar observations.  Existing mechanisms for producing such turbulence~\citep[e.g.][]{Ryu2008,Evoli2011} rely on structure formation and/or the galactic winds, and so can only produce IGM turbulence close to collapsed structures.  We explore in this Letter whether this can be achieved by turbulence driven by reionization \footnote{Galactic and AGN feedback might be the other possibility \citep[][]{Rieder2016}, although this is not anticipated to be volume filling.}.

This work is organized as follows.  In \S\ref{sec:numerical}, we describe our numerical setup, and, in \S\ref{sec:results}, we describe the dynamics of IGM turbulence.  We explore the possibility that turbulence could amplify magnetic fields in the IGM in~\S\ref{sec:Bfieldgrowth} and conclude in \S\ref{sec:conc}.  Throughout, we assume the following cosmological parameters: $\Omega_m = 0.305$, $\Omega_{\Lambda} = 1 - \Omega_m$, $\Omega_b = 0.048$, $h = 0.68$, $n_s = 0.9667$ and $\sigma_8 = 0.82$, consistent with~\citet{Planck2018} results. Distances are in co-moving units unless otherwise specified.  

\section{Numerical Methods}
\label{sec:numerical} 

The numerical methods used in this work are similar to those described in~\citet{DAloisio2020} and~\citet{Cain2024a}.  Here, we briefly describe the aspects of our methods relevant for this work, deferring further details to a forthcoming paper.  

Our simulations of small-scale IGM gas dynamics were run using a modified version of the RadHydro code of~\citet{Trac2004} and~\citet{Trac2007}.  Our version of RadHydro performs plane-parallel ray tracing radiative transfer (RT), with rays traced from the boundaries of cubical sub-volumes (RT domains) within our simulation box.   Gravity is solved using the particle-mesh (PM) method.  Our simulations are initialized at $z = 300$ using separate transfer functions for dark matter and baryons from CAMB~\citep{Lewis2000} and are run to $z = 4$.  Our fiducial box size is $L_{\rm box} = 250$ $h^{-1}$kpc, with
$1024^3$ gas cells, RT cells, and dark matter particles, for a fiducial spatial resolution of $\Delta x = 244$ $h^{-1}$pc.  Boundary conditions for the hydrodynamics and gravity are periodic~\footnote{Note that because of the construction of the RT domains (in which rays terminate that edges of domains), explicitly periodic boundary conditions for the RT is not necessary.  }.  

We run the simulation with hydrodynamics and gravity only (no RT) until the reionization redshift, $z_{\rm re}$.  Plane-parallel ionization fronts (I-fonts) cross the RT domains at a redshift $z_{\rm re}$, following the procedure described in~\citet{Cain2024a} (see also \S\ref{app:self_shielding} in supplemental material).  We use the reduced speed of light approximation to save computation time after I-fronts have crossed the RT domains.  Following~\citet{DAloisio2020}, the ionizing spectrum is discretized into $5$ frequency bins between $1$ and $4$~Ry, with a power law dependence $J_{\nu} \propto \nu^{-1.5}$.  The strength of the ionizing background is set such that the HI photo-ionization rate is $\Gamma_{\rm HI} = 3 \times 10^{-13}$ s$^{-1}$ everywhere (except within self-shielded systems).  To study numerical convergence, we have also run a simulation with $2048^3$ grid cells and dark matter particles ($\Delta x = 122$ $h^{-1}$pc), and ones with lower resolution~\footnote{ 
Due to computational limitations, we flash-ionized this $2048^3$ run instead of tracking RT (that is, we imposed a uniform ionizing background everywhere at $z = z_{\rm re}$).  We show in the supplemental material that this approximation does not significantly alter the subsequent relaxation.  }

\begin{figure*}
    \centering
    \vspace{-0.5cm}
    \includegraphics[scale=0.177]
    {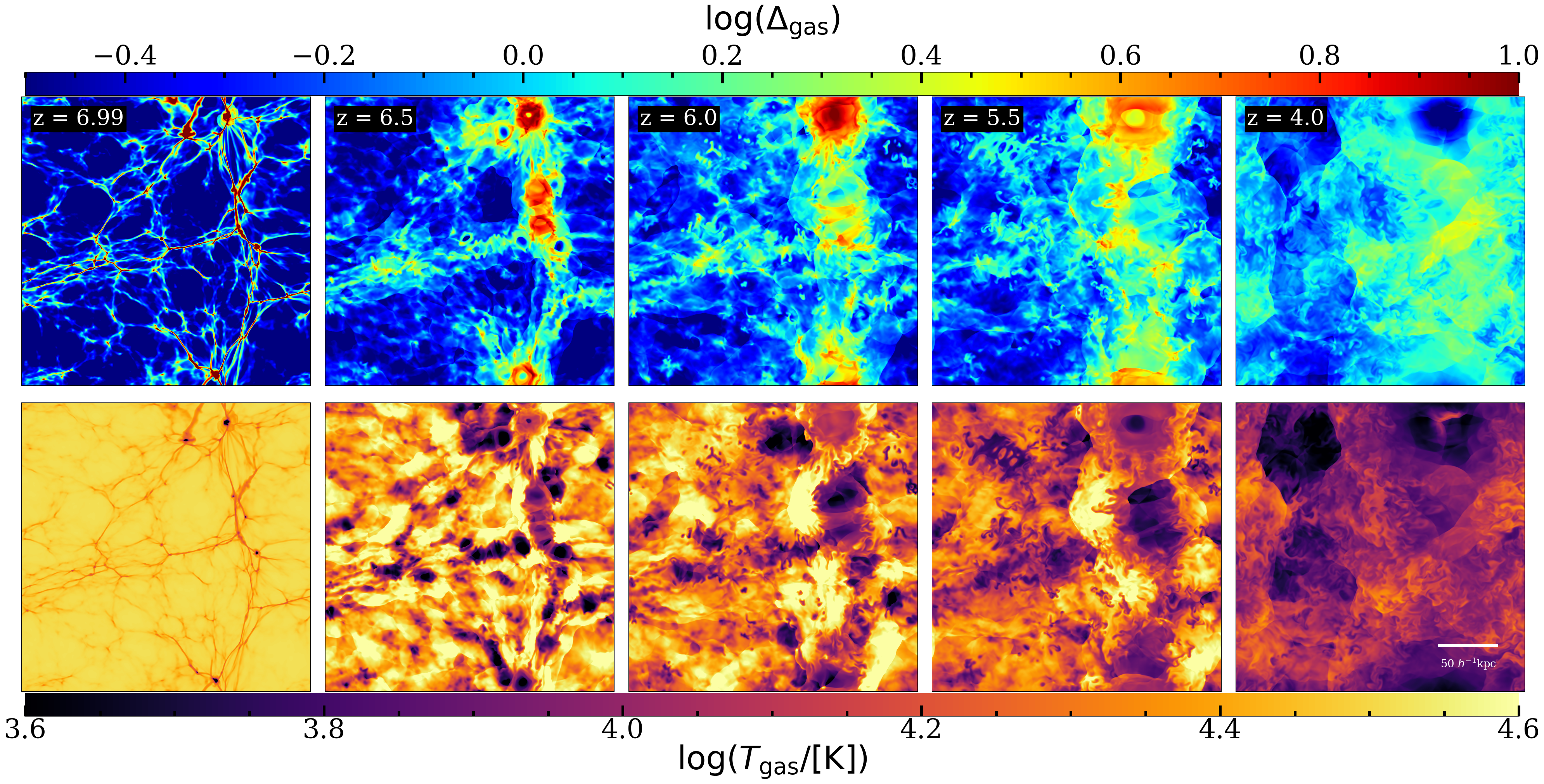}
        \vspace{-0.5cm}
    \caption{Snapshots showing the evolution of the density (top panels) and temperature (bottom panels) in our fiducial simulation.  Immediately after ionization ($z = 6.99$), the gas is clumpy at sub-kpc scales, reflecting its pre-ionized cold temperature, and is nearly isothermal, reflecting the heat input from reionization.  Subsequently, over-dense filaments expand, generating a medium of expansion-cooled and compression-heated gas.  By $z = 6$, the gas begins to show signs of turbulence, and at $z \leq 5.5$, turbulence is observed throughout, in both the density and temperature.  As relaxation slows, the turbulence dissipates, but some still persists to at least $z = 4$.  The inset in the bottom right shows the physical scale. }

    \label{fig:time_series_visualization}
\end{figure*}

\section{Small-scale IGM turbulence}
\label{sec:results}

\subsection{Dynamics}

Figure~\ref{fig:time_series_visualization} shows the time evolution of the IGM gas density (top panels) and temperature (bottom panels) in our fiducial simulation, where reionization occurs at a redshift of $z_{\rm re} = 7$.  At $z = 6.99$, the gas has had negligible time to respond to the photo-heating and retains its initial clumpiness.  The low-density gas is nearly isothermal, at $\approx 30,000~$K, while the over-dense filaments are slightly cooler.  These temperature differences result from the I-fronts moving more slowly through filaments, since the resulting temperature scales with the local I-front speed ~\citep{DAloisio2019,Zeng2021}. 

\begin{figure*}
    \centering
    \includegraphics[scale=0.43]{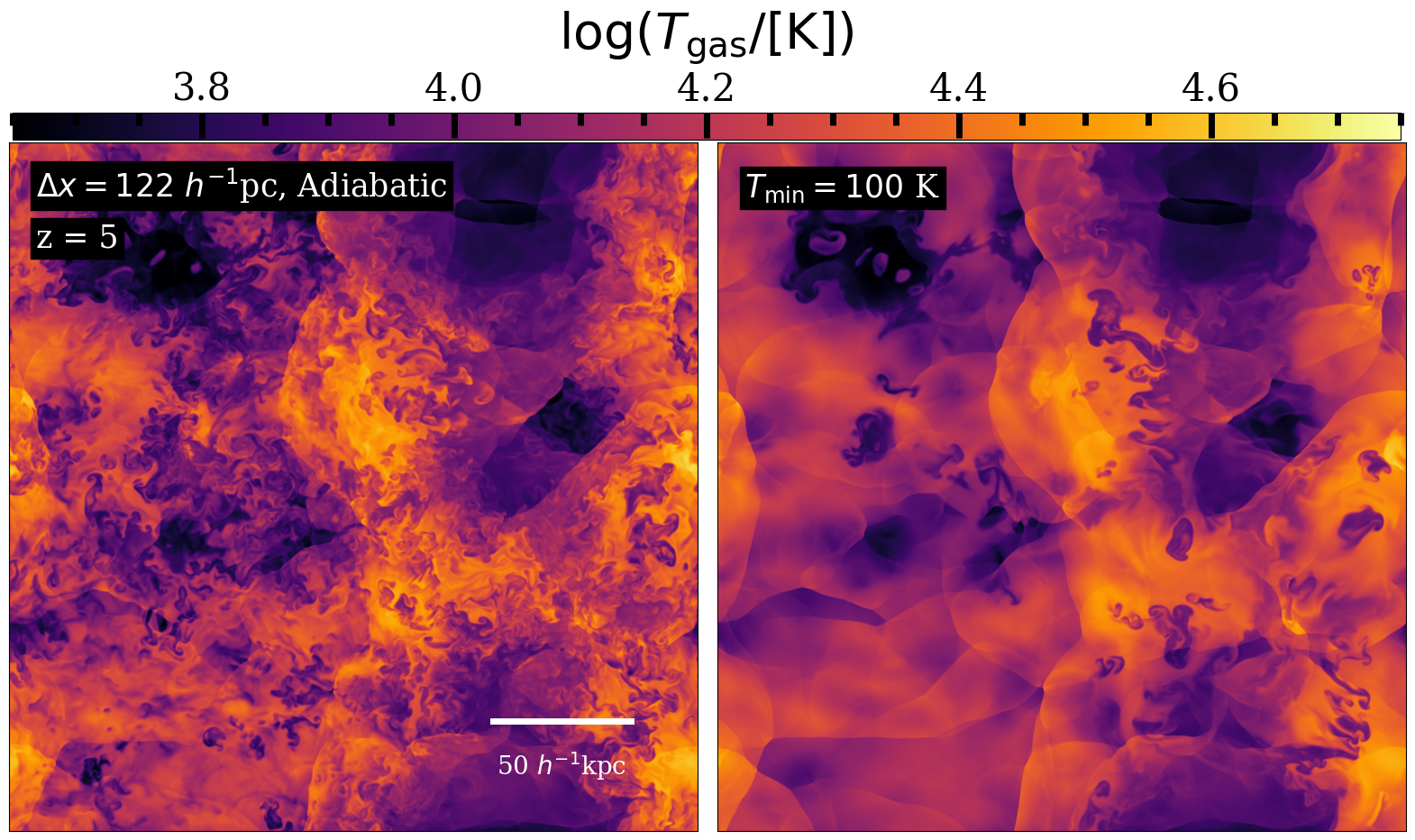}
    \caption{Comparison of turbulence in runs with and without X-ray pre-heating, showing our highest resolution $0.25~h^{-1}~$Mpc, $2048^3$ simulations.  {\bf Left}: slice through the temperature field at $z = 5$. The simulation assumes no heating before reionization such that the thermal evolution is essentially adiabatic.  {\bf Right}: the same, but for a run with a minimum gas temperature of $100$ K imposed at $z < 15$.  We see less turbulence in this case, with it being concentrated largely around the filaments.  }
    \label{fig:highres_Tmin1e2}
\end{figure*}

Between $z = 7$ and $6.5$, the gas dynamics are consistent with the findings of~\citet{Cain2024a} (see also~\citet{Hirata2018}).  The filamentary structures expand and cool adiabatically in response to impulsive heating from reionization.  Meanwhile, the gas surrounding the expanding filaments is compressed and heated to $\gtrsim 50,000~$K.  By $z = 6.5$, there is a factor of $\approx 10$ temperature difference between over-dense, expanding gas and surrounding hot, compressed gas at lower densities.  
At $z = 6$, these two phases are beginning to turbulently mix, and, by $z = 5.5$, turbulence has developed, and is clearly seen in the density and temperature fields.  Some of the turbulence may be instigated by Rayleigh-Taylor instabilities of less dense but over-pressurized gas expanding into denser but lower pressure regions.  Regardless, it is clear that the driving mechanism of the turbulence in these simulations is the pressure-smoothing of filaments and mini-halos and the resulting overlap of gas evacuating these structures.  By $z = 4$, $700$ Myr after $z_{\rm re},$ relaxation is nearly complete, and the turbulence has started to dissipate.  

No previous works (to our knowledge) have observed turbulent behavior in the context of the reionized, diffuse IGM. 
We believe this is because they did not achieve the required spatial resolution.  The highest-resolution studies to-date in this context (using grid or particle-based codes) achieve resolutions of $\approx$ $1-2$ ckpc~\citep{Park2016,Hirata2018,DAloisio2020,Chan2023,Cain2024a}.  The focus of these studies was resolving the pre-ionization Jeans scale -- the smallest scale at which the gas clumps --, which corresponds to $\approx 1-10$kpc (see Appendix A of~\citep{DAloisio2020}).  We have run a convergence test (see supplemental material) and found that turbulent features are not seen at all in our simulations for cell sizes larger than $0.5$ $h^{-1}$kpc.  We show in Appendix~\ref{app:analytical} that this resolution requirement is set not only by the Jeans scale of the ionized gas, but also by the scale at which numerical discretization introduces effective viscosity that prevents turbulence from forming~\citep{Teyssier2015}.  We further show that in the context of post-reionization pressure smoothing, resolutions of $0.5-1$ kpc are the expected minimum requirement to start resolving turbulence.

\subsection{Effect of X-ray pre-heating}
\label{subsec:xrays}

Pre-heating by X-rays before reionization increases the Jeans scale of the pre-ionized gas above that of the limit of pure ``adiabatic'' evolution that is assumed by our fiducial simulation~\footnote{Note that Compton heating off the CMB is significant at $z \gtrsim 100$, and we do include this effect.  At $z \lesssim 100$, cosmic expansion and X-ray pre-heating at the only significant heating/cooling channels before reionization. 
 }.  
Smoothing of small-scale structure by pre-heating could inhibit the formation of turbulence by eliminating the smallest structures and making it less likely for expanding filaments to overlap.  Upper limits on the 21 cm power spectrum from HERA~\citep{Hera2022} already constrain the temperature of the neutral IGM to be $\gtrsim 10$ K by $z \approx 8$.  Theoretical expectations for the minimum temperature, $T_{\min}$, of the pre-heated IGM are as high as $T_{\min} \approx 1000$K~\citep[e.g.][]{Furlanetto2006b}.  However, recent work presenting more careful modeling of the spectra of high-redshift X-ray binaries (the most plausible source of such heating) anticipates temperatures closer to $\approx 100$ K near the start of reionization~\citep{Fialkov2014c}.  
We model the effects of X-ray heating in an approximate manner, following~\citet{DAloisio2020}, by imposing a temperature floor of $T_{\min}$ at $z < 15$.    
Our pre-heating implementation likely overestimates the effect of pre-heating on structure formation for a given $T_{\min}$, since X-ray pre-heating is likely to trace the exponential growth of structure formation~\citep[e.g.][]{Davies2023b}.  

In Figure~\ref{fig:highres_Tmin1e2}, we show slices through the temperature at $z = 5$ for our highest-resolution adiabatic simulation with $122$ $h^{-1}$pc resolution (left), and the same run but with $T_{\min} = 100$ K (right).  As in Figure~\ref{fig:time_series_visualization}, we see turbulence throughout the box in the left panel.  On the right, however, turbulent features are much more sparse, favoring regions around the most massive filaments.  The sparsity of dense structures before reionization makes it less likely once reionized for their photo-evaporating remnants to collide, which suppresses the driving mechanism of the turbulence.  We note that the formation of massive structures, around which turbulence still forms in the $T_{\min} = 100$K run, is suppressed in our tiny boxes.   As such, the fraction of gas that is turbulent in our $T_{\min} = 100~$K run is likely an underestimate for the entire IGM.  
Appendix~\ref{app:quant} and the supplementary material (\S\ref{subsec:modeling}) presents a $T_{\min}=1000~$K simulation, in which there is no evidence for turbulence.

\section{The turbulent dynamo \& B-field growth}
\label{sec:Bfieldgrowth}

Turbulence will act to grow magnetic fields via the turbulent dynamo.  We show in Appendix~\ref{app:quant} that our simulations exhibit the Kolmogorov energy spectrum scaling expected for turbulent gas ($\propto k^{-5/3}$), down to a maximum $k$ set by the resolution of the simulation (see Appendix~\ref{app:analytical}).  In the absence of this ``numerical viscosity'', we expect to see turbulence extend down to a minimum scale set by the viscosity of intergalactic hydrogen, which works out to a co-moving scale of about a parsec for our estimated Reynolds number of ${\rm Re} \sim 10^5$ (Appendix~\ref{app:quant}). 

If a fraction $f_B$ of the turbulent kinetic energy goes into magnetic fields near the time of reionization, this would result in present-day $B$ fields of
\begin{eqnarray}
B &=&  \left(1 +z_{\rm re} \right)^{-2} \sqrt{4 \pi m_p n_e V_{\rm dr}^2 f_B}; \\
&=& 1.5\times 10^{-9} {\rm G} ~\times f_B^{1/2}\left( \frac{1+z_{\rm re}}{7} \right)^{-1/2} \left( \frac{V_{\rm dr}}{20 ~{\rm km ~s}^{-1}} \right),
\label{eqn:Beq}
\end{eqnarray}
where $m_p$ is the proton mass, $n_e$ the electron density, and $V_{\rm dr}$ is the velocity of the driving.  We have assumed the present-day field is reduced from that generated by $(1 +z_{\rm re})^{-2}$, as would occur from homogeneous cosmological expansion, and we note that the physical coherence length is also dilated by $1+z_{\rm re}$  by expansion.  There is substantial disagreement about the nature of magnetic fields generated by turbulence as summarized in \citet{2022JPlPh..88e1501S}. While simulations generally suggest that the field saturates with $f_B\sim 1$, some have argued that the $B$-field from turbulence saturates on scales near the driving scale of turbulence, with the magnetic field spectrum tracing the turbulence \citep{Xu2016}.  Others have argued that the scale should be much smaller \citep{Schekochihin2004}.   

The tightest constraints on IGM B-fields come from TeV blazars. 
\citet{Neronov2010} and \citet{2018ApJS..237...32A} find that the lack of detection of extended point sources requires $B_{\rm IGM} >3\times10^{-16} (\lambda/10 ~{\rm kpc})^{-1/2}$G for $B$-field coherence lengths of $\lambda \lesssim 10~$kpc in their conservative analysis where they assume each blazar has been active for only ten years; the constraints rule out $30\times$ larger $B_{\rm IGM}$ if the blazar is assumed to have its current TeV spectrum for $10^4$ years.  Similar constraints have also been placed from a lack of GeV reprocessed photons in the spectral energy distribution  \citep{2010Sci...328...73N, 2011A&A...529A.144T}.  If we evaluate equation~(\ref{eqn:Beq}) at $z_{\rm re}=6$ and $c_s=20~$km~s$^{-1}$, the conservative constraint in \citet{2018ApJS..237...32A} implies the $B$-field must have coherence of at least 
\begin{equation}
\lambda > 1\times10^4 f_B^{-1} f_V^{-1}~ {\rm km},
\label{eqn:lambda_blazars}
\end{equation}
to explain the blazar constraints, where $f_V$ is the volume filling fraction of the turbulence and, hence, magnetic fields.

Models for turbulent dynamo suggest that reionization-driven turbulence would result in magnetic fields that satisfy the constraint given by equation~(\ref{eqn:lambda_blazars}). 
In the accepted picture for the initial development of (non-helical) turbulent dynamo, the magnetic field grows until it comes into near equipartition with the kinetic energy of the eddies on the viscous scale so that the magnetic energy is $\approx V^2\, {\rm Re}^{-1/2}$.  At this point, the magnetic field stretching by the viscous eddies gets suppressed, leading to `Batchelor level' saturation \citep{1950RSPSA.201..405B, Rincon_2019}.  The growth during this stage happens on the short turnover timescale of the smallest hydrodynamic eddies ($L/[V ~{\rm Re}^{1/2}]$).  However, the scale of magnetic fluctuations is concentrated at the resistive scale when this saturation first happens (the scale over which magnetic diffusion is faster than the turnover time), which is given by $k_\eta = {\rm Pr}^{1/2} k_\nu$, where $\text{Pr}$ is the magnetic Prandtl number, and we find ${\rm Pr} \sim 10^{15}$ for our fiducial inputs.  If this were the end of the story, even volume-filling turbulence would result in present-day coherence lengths of $\lambda \sim 10^5$~km and $f_B \sim 10^{-2}$, comparable to what is required to explain the blazar constraint given by equation~(\ref{eqn:lambda_blazars}).  This would be true even in cosmic voids far from star forming halos, of which our tiny boxes are likely representative.

Hydrodynamic driven-turbulence simulations show that magnetic fields grow to energies above the Batchelor level saturation (closer to equipartition with the kinetic energy at the driving scale) and achieve larger coherence lengths than $k_\eta^{-1}$ \citep{Rincon_2019}.  However, the applicable regime of $1 \ll \text{Re} \ll \text{Pr}$, is challenging to simulate and so studies tend to combine both analytic arguments and numerical results to draw conclusions about the saturated state.  \citet{2022JPlPh..88e1501S} suggested that instabilities associated with the tearing of eddy-aligned structures result in the field's power spectrum peaking at $({\rm Re\times Pr})^{1/3}/L$, or roughly a length-scale of $10^{10}~$km for our fiducial inputs. With some numerical support, \citet{2007JFM...575..111Y} and \citet{2022JPlPh..88e1501S} conjectured that the magnetic energy power spectrum may scale to lower wavenumbers in a manner such that there is equal magnetic energy per log$k$, and, given sufficient time, extend down to $k \approx L^{-1}$.  
Any such redistribution of energy to larger scales would further act to deflect the Blazar TeV pairs. 

A common hypothesis that is supported by some simulations \citep{Haugen_2003} is that turbulence saturates with the magnetic field having the same energy distribution as the turbulent kinetic energy \citep[e.g.][]{Rincon_2019, 2022JPlPh..88e1501S}.  
\citet{Xu2016} developed a quantitative theory for how such saturation would occur. 
Using their equations, we find that if there is $\approx 1$ turnover time at the driving scale as our simulations suggest (Appendix~\ref{app:quant}), the field would have cascaded to match the turbulent kinetic energy down to $k \approx 10~ L^{-1}$, which would be easily sufficient to quench the blazar pair halos.  We note, however, that precisely evaluating the growth of the magnetic field in such a scenario would require MHD simulations, since magnetic fields with energy comparable to the kinetic energy of the gas can modify the turbulent dynamics at any scale, if not non-ideal MHD simulations that resolve down to the kilometer scales over which magnetic fields diffuse.

When the turbulence is not volume filling, as we find occurs at $T_{\rm min} \gtrsim 100~$K, there is an additional condition that must be satisfied to deflect the pairs from TeV blazars and hence suppress their GeV halos. Namely, each $e^{\pm}$ pair should intersect over its path a region where there has been turbulent amplification of $B_{\rm IGM}$. At $z=0$, each pair travels $\approx 400 (E_e/1~\text{TeV})^{1/2}$~kpc before losing its energy by Compton scattering the CMB, where $E_e$ is the energy of one of the pairs. 
The typical energy of the up-scattered CMB photons is then $2 ~\text{GeV} (E_e/1 ~\text{TeV})^2$, and constraints that find missing energy are centered at $\approx 1~$GeV, extending in some cases to $\approx 10~$GeV \citep[e.g.][]{2011A&A...529A.144T}.  Such photons arise from $0.5-3~$TeV pairs and, hence, these particles travel $200 -800~$kpc.   The $250~$kpc$/h$ slices shown in Figure~\ref{fig:highres_Tmin1e2} illustrate that for $T_{\rm min} = 100~$K, even a $200~$kpc path would likely intercept a turbulent domain in which magnetic fields had been amplified. The existence of turbulence in the $T_{\rm min} = 100$ K case suggests that most regions in the IGM would still see some turbulence in this scenario.

\section{Conclusions}
\label{sec:conc}

For the first time, we have shown that the destruction of mini-halos and IGM filaments by reionization drives turbulent eddies in the IGM that may fill a significant fraction of the IGM volume at $z > 4$.  
Our main findings are summarized below: 

\begin{itemize}

    \item In the aftermath of passing ionization fronts, filaments and mini-halos are heated to $\approx 20,000-30,000$K and expand rapidly, cooling in their centers and heating up around their boundaries as rarefied gas is compressed. As these structures overlap, expansion-cooled and compression-heated gas begin to mix and form turbulent eddies.  These eddies are powered by the expansion of filaments until the IGM reaches a new pressure equilibrium after several hundred Myr.   

    \item  X-ray pre-heating eliminates small-scale structure prior to reionization, making it less likely for expanding filaments to overlap and generate turbulence. We find turbulence is somewhat suppressed by preheating to temperatures of  $T \approx 100~$K  and eliminated for temperatures of $T \approx 1000~$K.  Such turbulence may be patchy or even absent depending on the relative timing of pre-heating and reionization.  

    \item We showed that such turbulence would likely generate magnetic fields that could explain the lack of GeV halos towards TeV blazars, at least for $X$-ray preheating temperatures of $T \lesssim 100~$K.   This conclusion is likely robust to current debates regarding the character of the magnetic fields that would be generated by turbulence.  Reionization driving turbulence represents a plausible mechanism to amplify volume-filling magnetic fields throughout the IGM.  
    
\end{itemize}

If reionization-driven turbulence is the source of IGM magnetic fields, future $\gamma$-ray observations, utilizing also echoes and off-axis blazars, may be able to differentiate reionization-turbulence dynamo pictures \citep[e.g.][]{2021Univ....7..223A}.

\begin{acknowledgments}
The authors thank Simon Foreman for sharing his dedicated computational resources on the Sol supercomputer at Arizona State University.  Some computations for this work were run using computational resources distributed under NSF ACCESS allocations TG-PHY230158 and TG-PHY240332 and by the Sol supercomputer at Arizona State University.  CC acknowledges support from the Beus Center for Cosmic Foundations.  MM acknowledges support from NSF grant AST-2007012 and NASA grant 80NSSC24K1220. ES acknowledges support from NASA  Grant 80NSSC22K1265. AD acknowledges support from NSF grant AST-2045600. HT acknowledges support from NASA grant 80NSSC22K0821.

\end{acknowledgments}


\appendix

\section{Requirements for IGM turbulence}
\label{app:analytical}

\subsection{Spatial resolution}
\label{subsec:spatial}

The viscosity of diffuse, ionized hydrogen is very small~\citep[][Reynolds number $\sim 10^5$ for kpc-scale driving at the sound speed]{Spitzer1941}, such that flows are likely to be turbulent if a driving force is present.  Shortly after reionization, pressure imbalances are everywhere as the IGM relaxes, providing such a force.  Here, we analytically estimate the minimum spatial resolution required to capture this behavior, and ask whether it is consistent with our findings.  The maximum characteristic length scale of IGM turbulence is set by the distance, $\Delta x^{\rm turb}$, that gas can travel in the time since it was reionized, $\Delta t$.  Assuming the gas speed in the turbulent eddies is comparable to the sound speed in ionized gas, $c_s$, then this scale is
\begin{equation}
    \label{eq:Deltax_turb}
    \Delta x^{\rm turb} \approx 10 ~h^{-1}\text{kpc}~\left(\frac{c_s}{20 \,\text{km/s}}\right)\left(\frac{1+z}{7}\right)\left(\frac{\Delta t}{100 \,\text{Myr}}\right).
\end{equation}
where $20$ km/s is taken as a typical reference value for $c_s$.  The limiting factor in capturing turbulence is ``numerical viscosity'' set by the simulation resolution, $\Delta x_{\rm cell}$.  \citet{Teyssier2015} estimated the associated ``numerical Reynolds number'', ${\rm Re}_{\rm num},$ to be $\approx 2 L/\Delta x$ for their grid-based hydrodynamics simulations, where $L$ is the length scale associated with the turbulence (their Eq. 35, see also~\citet{Rieder2017}).  Taking their estimate and replacing $L$ with $\Delta x^{\rm turb}$ yields
\begin{equation}
    \label{eq:Renum}
    {\rm Re}_{\rm num} \approx \frac{2 \Delta x^{\rm turb}}{\Delta x_{\rm cell}}.
\end{equation}
Plugging in for $\Delta x^{\rm turb}$ and re-arranging gives the maximum cell size that can achieve a given ${\rm Re}_{\rm num}$ at the driving scale, 
\begin{equation}
    \label{eq:resolution}
    \Delta x_{\rm cell}^{\rm max} \approx 0.4 ~h^{-1}\text{kpc}~\left(\frac{c_s}{20 \,\text{km/s}}\right)\left(\frac{{\rm Re}_{\rm num}^{\min}}{50}\right)^{-1}\left(\frac{1+z}{7}\right)\left(\frac{\Delta t}{100 \,\text{Myr}}\right),
\end{equation}
where ${\rm Re}_{\rm num}^{\min} \approx 50$ is a typical minimum Reynolds number required to observe turbulent behavior~\citep{Brandenburg2005}.  Note that plugging in an $L$ smaller than the driving scale would require smaller $\Delta x_{\rm cell}$.  If the driving force of the turbulence is pressure-smoothing, and the time-scale for the IGM to reach pressure equilibrium is $\approx 300$ Myr~\citep{DAloisio2020}, then the maximum cell size that will allow for {\it any} turbulent behavior to emerge before pressure-smoothing is complete is
\begin{equation}
    \label{eq:min_resolution}
    \Delta x_{\rm cell}^{\rm max} (300\,{\rm Myr}) \approx 1.2 ~h^{-1}\text{kpc}~\left(\frac{c_s}{20 \,\text{km/s}}\right)\left(\frac{{\rm Re}_{\rm num}^{\min}}{50}\right)^{-1}\left(\frac{1+z}{7}\right).
\end{equation}
Our estimated maximum cell size of $\approx 1$ $h^{-1}$ckpc is roughly the highest spatial resolution achieved in dedicated studies of the diffuse, low-density IGM to date~\citep{DAloisio2020,Chan2023}, with most having resolution much coarser than this~\citep[][and references therein]{Doughty2023}.  It is little surprise, then, that such turbulence has not yet been previously observed in IGM simulations.  

\subsection{IGM pre-heating requirement}
\label{subsec:Jeans}

A key finding in this work, however, is that achieving the spatial resolution requirements above is not a sufficient condition for turbulence to emerge.  Indeed, our X-ray pre-heating results suggest there must be pre-existing small-scale structure in the IGM at or below the driving scale in order for pressure imbalances to give rise to turbulence.  That is, we found that the overlap of expanding filaments generates the needed conditions for turbulence to emerge.  The IGM Jeans scale gives a rough estimate of the characteristic sizes of collapsed structures present in the IGM at a given density and temperature, and also a rough estimate of the characteristic distance between such structures.  The co-moving Jeans scale is given by 
\begin{equation}
    \label{eq:Jeans}
    L_{\rm J} = 8.5 ~h^{-1}{\rm kpc} ~~\left(\frac{1+z}{7}\right)^{-1/2}\left(\frac{T}{10^2 \,{\rm K}}\right)^{1/2} {\Delta}^{-1/2},
\end{equation}
where $\Delta$ is gas density in units of the mean and $T$ is the gas temperature.  We can get a rough idea of the maximum temperature that will admit turbulence by requiring the pre-ionized Jeans scale to be smaller than the driving scale -- that is $L_{\rm J} < \Delta x^{\rm turb}$.  This can be re-expressed as a condition on the temperature:
\begin{equation}
    \label{eq:Tcondition}
    T \lesssim 1500~{\rm K}~~\left(\frac{c_s}{20 \,\text{km/s}}\right)^2  \left(\frac{1+z}{7}\right)^3 \left(\frac{\Delta t}{100 \,\text{Myr}}\right)^2 \left(\frac{\Delta}{10}\right),
\end{equation}
where we have taken densities in units of the cosmic mean of $\Delta \approx 10$ as the typical (initial) density of a structure responsible for generating turbulence in the diffuse IGM.  Eq.~\ref{eq:Tcondition} is consistent with our finding that turbulence disappears for $T_{\min} = 1000~$K (see supplementary material), since structures with $\Delta \gg 10$ are rare in our tiny volumes.  It also explains why turbulence only forms around the most massive structures in our $T_{\min} = 100~$K run, since the above condition becomes $T < 150~$K for $\Delta = 1$ gas.  Eq.~\ref{eq:Tcondition} confirms our numerical results and suggests that the emergence of small-scale turbulence should be sensitive to the level and timing of pre-heating before reionization.  

\section{Quantifying turbulence}
\label{app:quant}

In this appendix, we quantify the properties of turbulence in our simulations.  The turnover timescale at wavenumber $k = 2 \pi/L$ in a turbulent flow is 
\begin{equation}
    \label{eq:tau_eddy}
    \tau_{\rm eddy}(k) \approx \frac{L}{\sigma_{\rm v}(k)} = \frac{2\pi/k}{ [2kE(k)]^{1/2}},
\end{equation}
where $\sigma_{\rm v}$ is the standard deviation of the velocity field at the scale $L$ and 
\begin{equation}
\label{eq:Ek}
E(k) \equiv \sum_i\Delta_{\rm v_i}^2(k)/(2 k)
\end{equation}
is the so-called ``energy spectrum'', where $\Delta_{\rm v_i}^2$ is the dimensionless power spectrum of the $i^{\rm th}$ velocity component.  Many studies have shown that the Kolmogorov scaling of $E(k)\propto k^{-5/3}$ tends to be followed over the turbulent inertial range for isotropic, homogeneous turbulence~\citep[e.g.][]{Beattie2024}.  The total number of turnover times elapsed at scale $k$, $N_{\rm \tau}$, is given by integrating $\tau_{\rm eddy}^{-1}$ over time, starting from the time a sound wave can cross the distance scale $2\pi/k$ since the start of reionization.  

\begin{figure}
    \centering
    \includegraphics[scale=0.28]{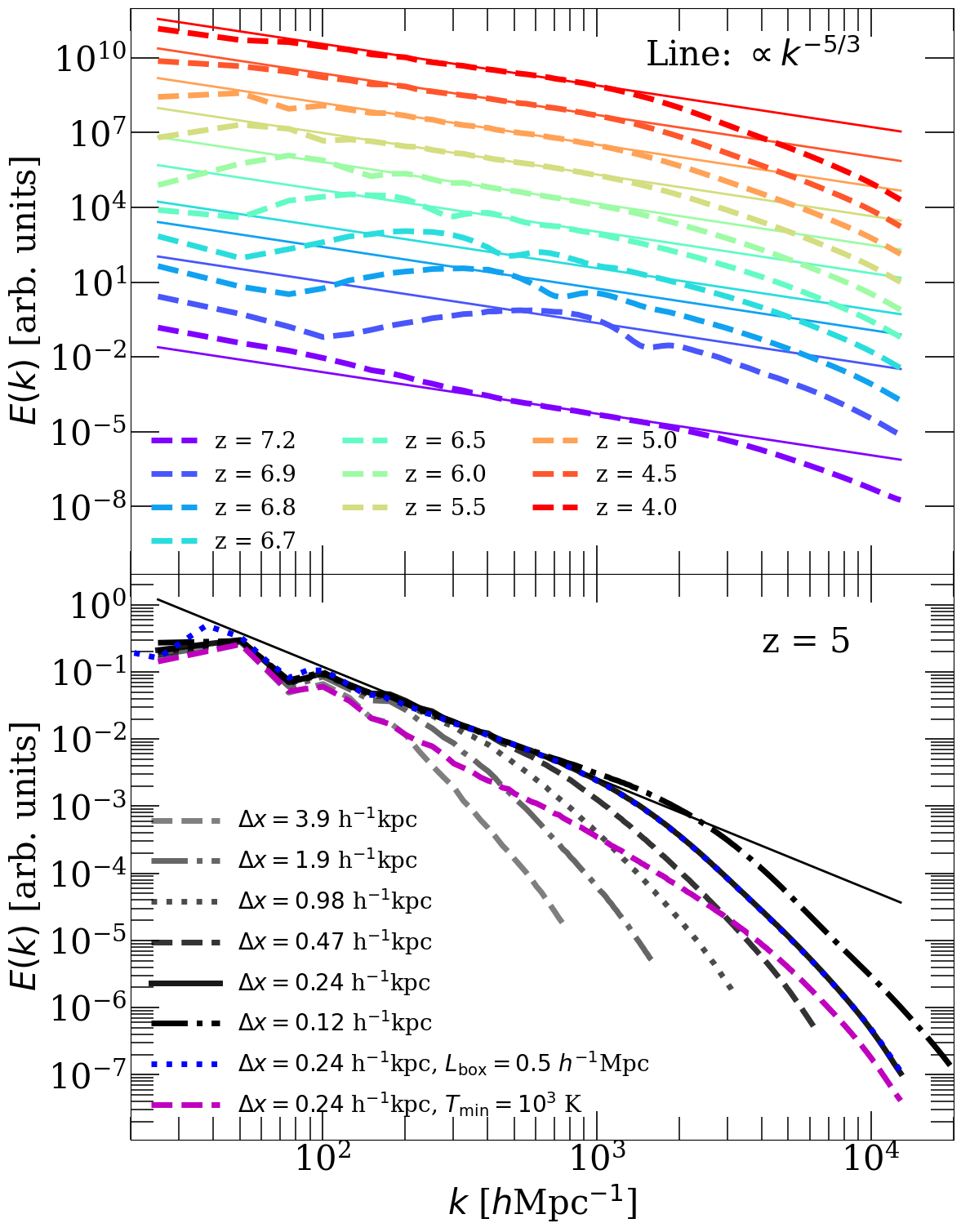}
    \caption{Quantification of turbulence in our simulations.  {\bf Top:} The energy spectrum $E(k)$ at a time-series of redshifts, offset vertically for visualization, with redshift decreasing from bottom to top.  The ``bump'' that moves to the left with time tracks the growing sound crossing distance since $z_{\rm re}$.  At smaller scales, we find the $k^{-5/3}$ scaling characteristic of turbulence, with a fall-off at very small scales due to numerical dissipation.  {\bf Bottom:} the same as the top panel except without rescaling the amplitude and concentrating on $z = 5$ and different spatial resolutions (gray curves).  The high-$k$ fall off shifts to the right as resolution increases, as expected if this cutoff is numerical.  The scaling at smaller $k$ is close to, but slightly shallower than, the $-5/3$ Kolmogorov scaling.  The blue dotted line is a $0.5$ $h^{-1}$Mpc box with the fiducial resolution, and shows convergence of $E(k)$ with box size.  The magenta-dashed curve shows a scenario with high X-ray pre-heating ($T_{\min} = 1000$ K), for which we do not observe the $-5/3$ scaling.  }
    \label{fig:ke_scaling}
\end{figure}

In the top panel of Figure~\ref{fig:ke_scaling}, we show $E(k)$ at a number of redshifts (decreasing from bottom to top) for our fiducial simulation.  We use an arbitrary vertical scaling and offset the curves in sequence to easily compare their shapes.  We offset the curves vertically so they can be easily compared, and show the expected $k^{-5/3}$ Kolmogorov scaling as a thin line.  At $z = 7.2$, just before ionization, $E(k)$ is slightly steeper than Kolmogorov, indicating that the dynamics are not dominated by turbulence.  A small amount of turbulence may be present in the densest structures due to non-linear collapse~\citep{Ryu2008}, but if this is present it fills a small fraction of the volume.  Just after ionization at $z = 6.9$, $E(k)$ has a prominent ``bump'' centered at $\approx 800$~$h~$Mpc$^{-1}$.  The right edge of the bump at $\approx 1500$~$h$~Mpc$^{-1}$ corresponds to a co-moving distance scale of $\approx 4$ $h^{-1}$kpc, roughly twice the distance that a $20$~km~s$^{-1}$ sound wave can travel through a $T \approx 2 \times 10^4$K highly photo-ionized gas in the $15$~Myr since $z_{\rm re} = 7$.  It is thus reasonable to interpret the bump as tracking the sound crossing distance since reionization.  At successively lower redshifts, the bump shifts to smaller $k$ as sound waves travel further from their source, disappearing by $z \approx 5$ as the sound crossing distance approaches the box scale.  

We expect turbulence to form at scales smaller than the sound-crossing distance -- that is, to the right of the bump.  This is exactly what we see -- starting at $z = 6.5$, $E(k)$ reaches the $k^{-5/3}$ scaling at a small range of $k$ around $600$ $h$Mpc$^{-1}$.  At lower redshifts, the $k$ range where this scaling holds increases, reaching as low as $\approx 200$ $h$Mpc$^{-1}$ by $z = 4$.  The evolution of $E(k)$ with time quantitatively confirms our earlier assertion that the pressure-smoothing of small-scale structure is the driving mechanism of turbulence in our simulations, as opposed to structure formation~\citep{Ryu2008} or some other mechanism.  The snapshots before and close to $z_{\rm re}$ do not show a spectrum with a shape characteristic of turbulent motions, and this behavior only emerges $\sim 200$ Myr after $z_{\rm re}$.  This time delay is consistent with the timescale required for expanding filaments to begin overlapping.  

In the bottom panel, we show $E(k)$ at $z = 5$ for simulations with varying spatial resolution (black/gray curves).  The scaling at turbulent scales is slightly shallower than $k^{-5/3}$ -- the green dot-dashed curve is best fit by a $k^{-1.45}$ scaling.  This may be because turbulent eddies do not fill the entire volume, that the scale of turbulent driving $L\approx c_s \Delta t$ is increasing with time, and/or that some non-turbulent kinetic motion is present.  The fall-off in $E(k)$ shifts to the right with increasing resolution, confirming that this is due to numerical dissipation.  We also show $E(k)$ for a simulation with box size $L = 0.5$ $h^{-1}$cMpc (blue dotted) and $T_{\rm min} = 10^3$K (magenta dashed).  The first demonstrates that the energy spectrum at turbulent scales is converged with respect to box size, and the second shows that extreme X-ray pre-heating erases turbulence (the magenta curve does not follow the $k^{-5/3}$ scaling), as qualitatively observed in Figure~\ref{fig:highres_Tmin1e2}.  

In reality, when the turbulence is hydrodynamic the cascade should extend to $k_\nu \approx {\rm Re}^{3/4}/L$ where the Reynolds number for mean density gas is given by
\begin{eqnarray}
{\rm Re} &\equiv& \frac{ V_{\rm dr} L}{\nu} = 2\times10^5 \left( \frac{V_{\rm dr}}{20 ~{\rm km ~s}^{-1}}  \right) \left( \frac{L}{1 ~{\rm kpc}}  \right) \nonumber \\
&& \times \left ( \frac{1+z}{7} \right)^3 \left( \frac{T}{2\times10^4 {\rm K}}  \right)^{-5/2},
\end{eqnarray}
where $\nu$ is the kinematic viscosity for mean-density ionized gas.  This indicates that the hydrodynamic turbulence can extend to sub-parsec scales for kpc driving and applicable parameters. As discussed in the next section, the short turnover time of these smaller eddies (with the smallest eddies turning over a time $\text{Re}^{-1/2}$ times shorter than the largest) allows turbulence to grow magnetic fields even though we find that the smallest eddies we can resolve turnover only $N_{\rm \tau} \approx 1$ time by $z = 4$.  We find that $N_{\tau}$ is smaller than this when pre-heating by X-rays is significant (as low as $\approx 0.1$ for $T_{\min} = 1000$ K, see Supplemental Material for more detail).

\bibliography{./references}

\clearpage

\section{Supplemental Material}

\subsection{Conditions for achieving turbulence}

\subsubsection{Spatial resolution}
\label{subsec:convergence}

\begin{figure*}
    \centering
    \includegraphics[scale=0.15]{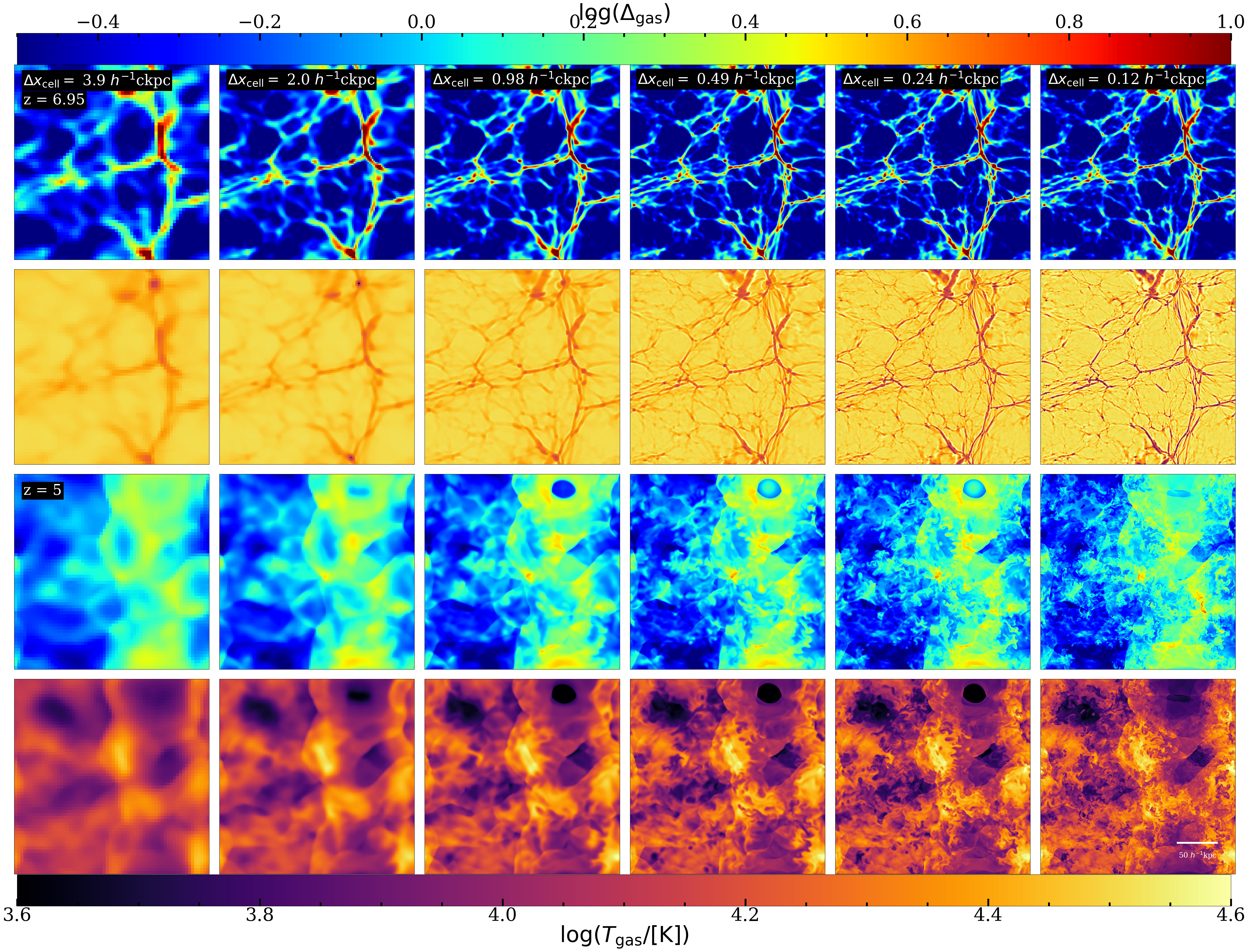}
    \caption{Convergence of small-scale IGM turbulence with spatial resolution.  {\bf Top two rows}: density and temperature at $z = 6.95$ for $N_{\rm RT} = 64^3$, $128^3$, $256^3$, $512^3$, $1024^3$ (our fiducial resolution), and $2048^3$ (flash ionized).  The amount of small-scale structure in the density and temperature fields just after ionization noticeably increases from left to right, and it is not well-converged until our fiducial resolution.  {\bf Bottom two rows}: the same, but at $z = 5$ when IGM turbulence has become significant in our fiducial simulations.  The lowest resolution at which turbulent structures are clearly seen is $N_{\rm RT} = 512^3$, and the turbulent structure is not converged even at $N_{\rm gas} = 2048^3$.  For relevant IGM parameters, turbulent eddies are expected to be present two orders of magnitude below this spatial resolution.  }
    \label{fig:convergence_visualization}
\end{figure*}

We consider the convergence with spatial resolution of the turbulent effects seen in Figure~\ref{fig:time_series_visualization}.  In Figure~\ref{fig:convergence_visualization}, we show density and temperature maps at $z = 6.95$ (top two rows) and $z = 5$ (bottom two rows) for $N = 64^3$, $128^3$, $256^3$, $512^3$, $1024^3$, and $2048^3$ (columns from left to right).  Note that the $N = 2048^3$ run is flash-ionized and does not include RT.  In the top rows, we see that the amount of small-scale structure in the density and temperature maps just after ionization increases from left to right.  The density and thermal structure of the gas just after ionization converges reasonably well at our fiducial resolution, $0.24$ $h^{-1}$kpc, but at coarser resolutions some small-scale structure is clearly missing.

In the bottom panels, we see that the turbulent behavior in the pressure smoothed IGM is not converged, even at the highest resolution we simulate of $\Delta x = 0.12$ $h^{-1}$kpc.  Indeed, we do not even see turbulent behavior emerge at all for resolutions coarser than $0.5$ $h^{-1}$kpc, consistent with the arguments in \S\ref{subsec:spatial}.  An interesting finding is that even though the pre-ionization structure of the gas seems reasonably well-converged at our fiducial resolution, the turbulent features at $z = 5$ are clearly not.  This suggests a non-trivial relationship between the numerical requirements for resolving the structures that seed turbulence and the conditions to resolve the turbulence itself (see Appendix~\ref{app:analytical} in the main text).  

\begin{figure*}
    \includegraphics[scale=0.44]{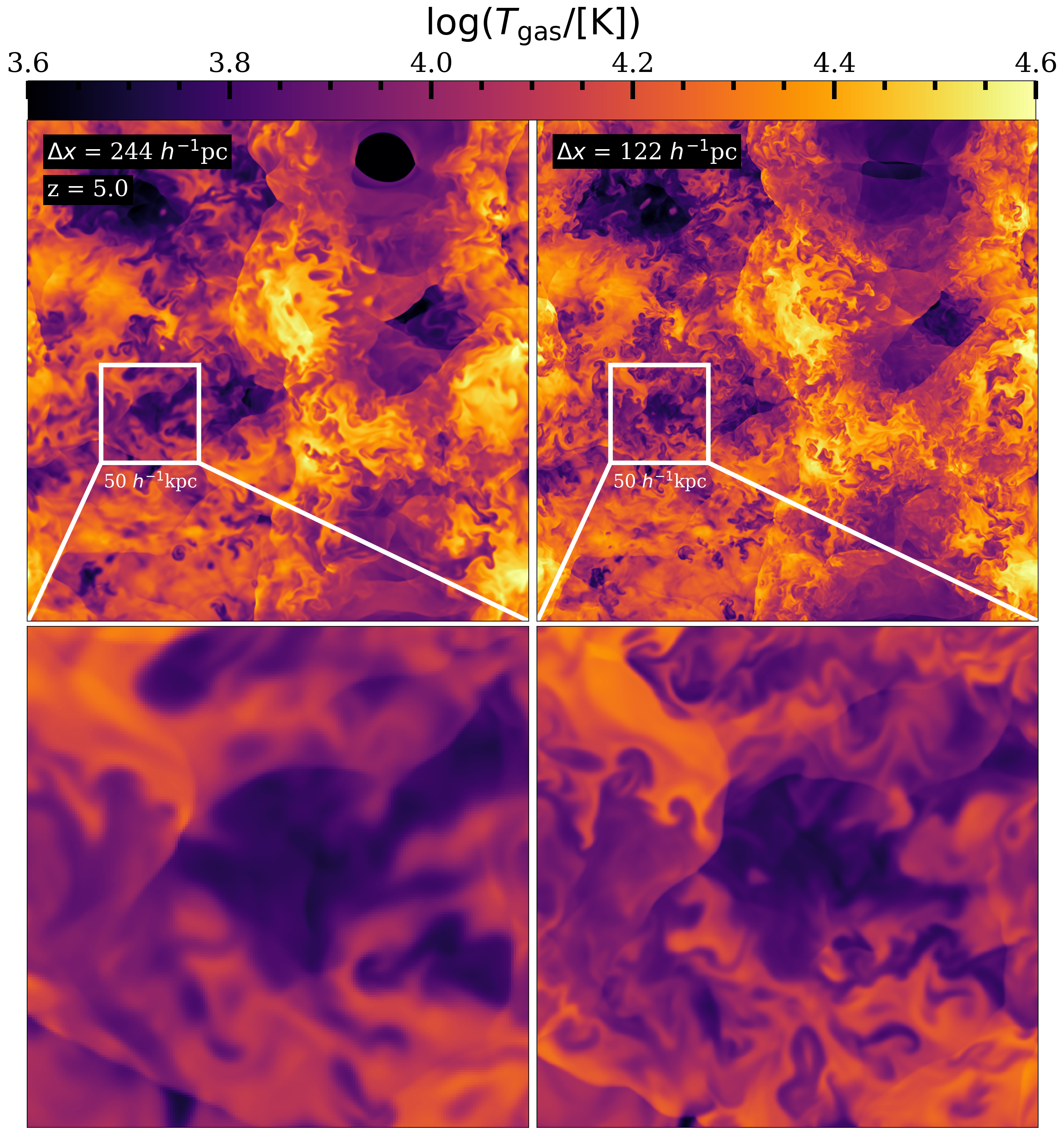}
    \caption{Comparison of the temperature slice at $z = 5$ in Figure~\ref{fig:time_series_visualization} at our fiducial resolution of $\Delta x = 244$ $h^{-1}$pc, and the same slice in a flash-ionized box with $\Delta x = 122$ $h^{-1}$pc.  The top panels show the same temperature snapshots as in the lower right of Figure~\ref{fig:convergence_visualization}, and the bottom panels show a $50$ $h^{-1}$kpc zoom-in.  Turbulent features are much more prominent in the higher-resolution run and are anticipated to continue down to $\sim 1~$pc. 
    }
    \label{fig:z6.25_zoom}
\end{figure*}

In Figure~\ref{fig:z6.25_zoom}, we take a closer look at the $z = 5$ temperature snapshots from our two highest-resolution runs. The top panels compare the same snapshots shown in the lower right of Figure~\ref{fig:convergence_visualization}, and the bottom panels show a zoom-in on a $50$ $h^{-1}$kpc region.  We see that in both cases, the turbulent features fill nearly the entire volume, and are missing only in cold pockets where gas previously escaped from the deepest potential wells.  The turbulent eddies are much more prominent in the highest-resolution run, as the zoom-in clearly shows. Turbulent features are anticipated to continue down to $\approx 1~$pc, far below the $\approx 100$ pc resolution in our highest resolution simulations.    

\subsubsection{Box size \& local density}
\label{subsec:box_size}

\begin{figure*}
    \centering
    \includegraphics[scale=0.6]{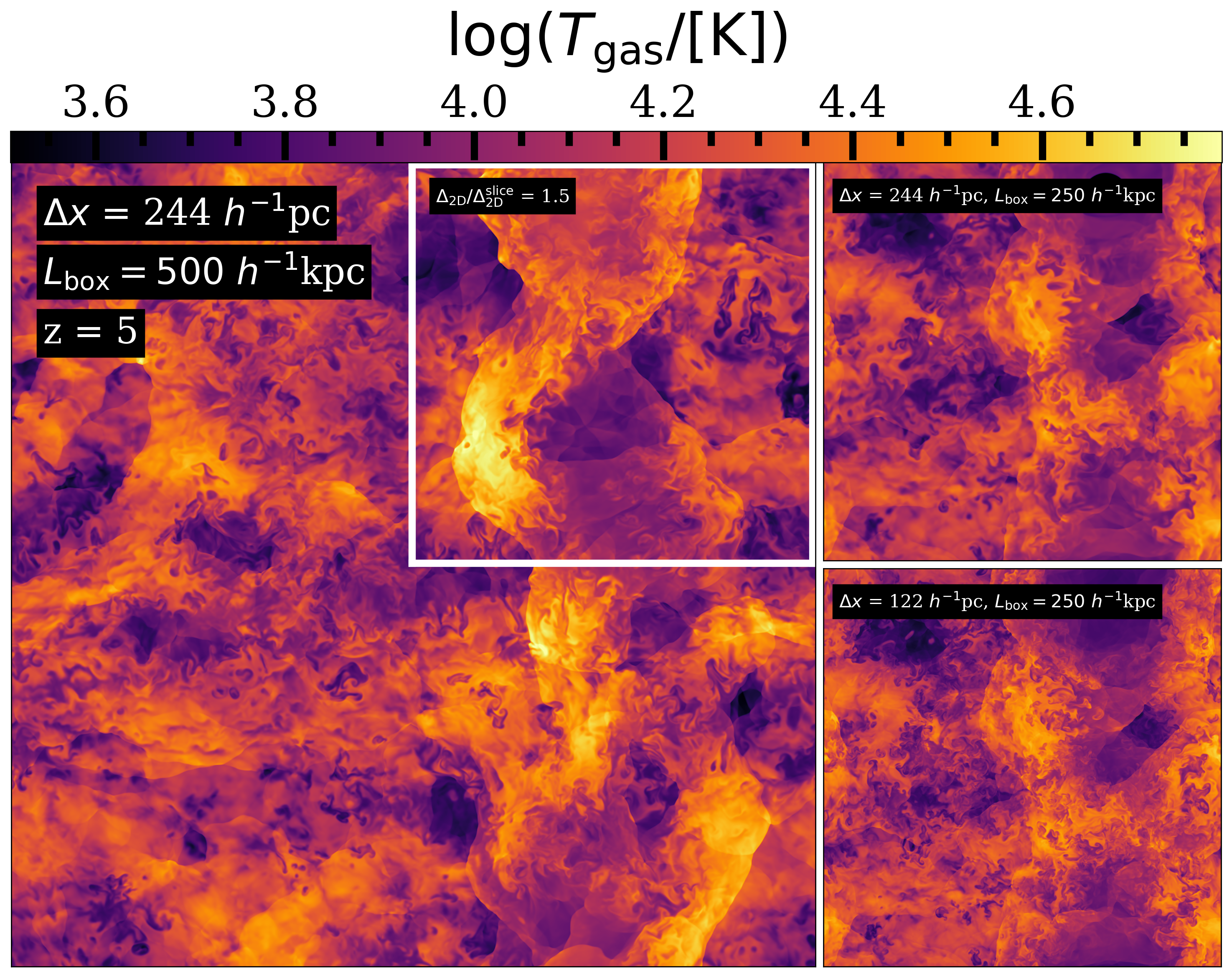}
    \caption{Visualization of the effect of box size on IGM turbulence.  On the left, we show a slice through the temperature in our $0.5$ $h^{-1}$Mpc, flash-ionized box at $z = 5$.  The upper-right quadrant, which is highlighted, is mildly over-dense relative to the rest of the slice.  The small panels on the left are the same slices shown in Figure~\ref{fig:z6.25_zoom}, re-sized to scale.  We see no clear qualitative differences in turbulent features as a function of box-size or local over-density.  }
    \label{fig:L05_box_vis}
\end{figure*}

Our fiducial simulations are run in extremely small boxes relative to typical cosmological simulations, sampling a patch of the IGM at the cosmic mean density.  To study the effect of box size on our results, we ran an $L = 0.5$ $h^{-1}$Mpc, flash-ionized box with $N = 2048^3$, matching our fiducial cell size of $0.244$ $h^{-1}$kpc.  We show a slice through the temperature field in this run at $z = 5$ in Figure~\ref{fig:L05_box_vis}, alongside the same for the two smaller boxes shown in Figure~\ref{fig:z6.25_zoom}, re-sized to scale.  The upper-right quadrant of the $0.5$ $h^{-1}$Mpc slice, which we highlight, is mildly over-dense (a factor of $1.5$ relative to the rest of the slice).  This sub-volume is dominated by a large filamentary structure that was recently destroyed by reionization.  

Visually, there is no clear indication that the turbulent features are qualitatively different in boxes with different sizes.  They appear to form at the same scale as in our fiducial box, and to fill roughly the same fraction of the slice.  One might expect the destruction of more massive filaments to drive stronger turbulence due to the larger gas pressures driving the expansion.  However, comparing the boxed region on the left to the right panels, there is no clear indication that this is the case.  This suggests that our fiducial box size is sufficiently large to capture the essential properties of the turbulence.  However, it is unclear whether {\it much} larger volumes, containing halos that would host star-forming galaxies, would have much different properties in over-densities hosting such halos.  

\subsubsection{Vorticity}
\label{subsec:vorticity}

\begin{figure*}
    \centering
    \includegraphics[scale=0.45]{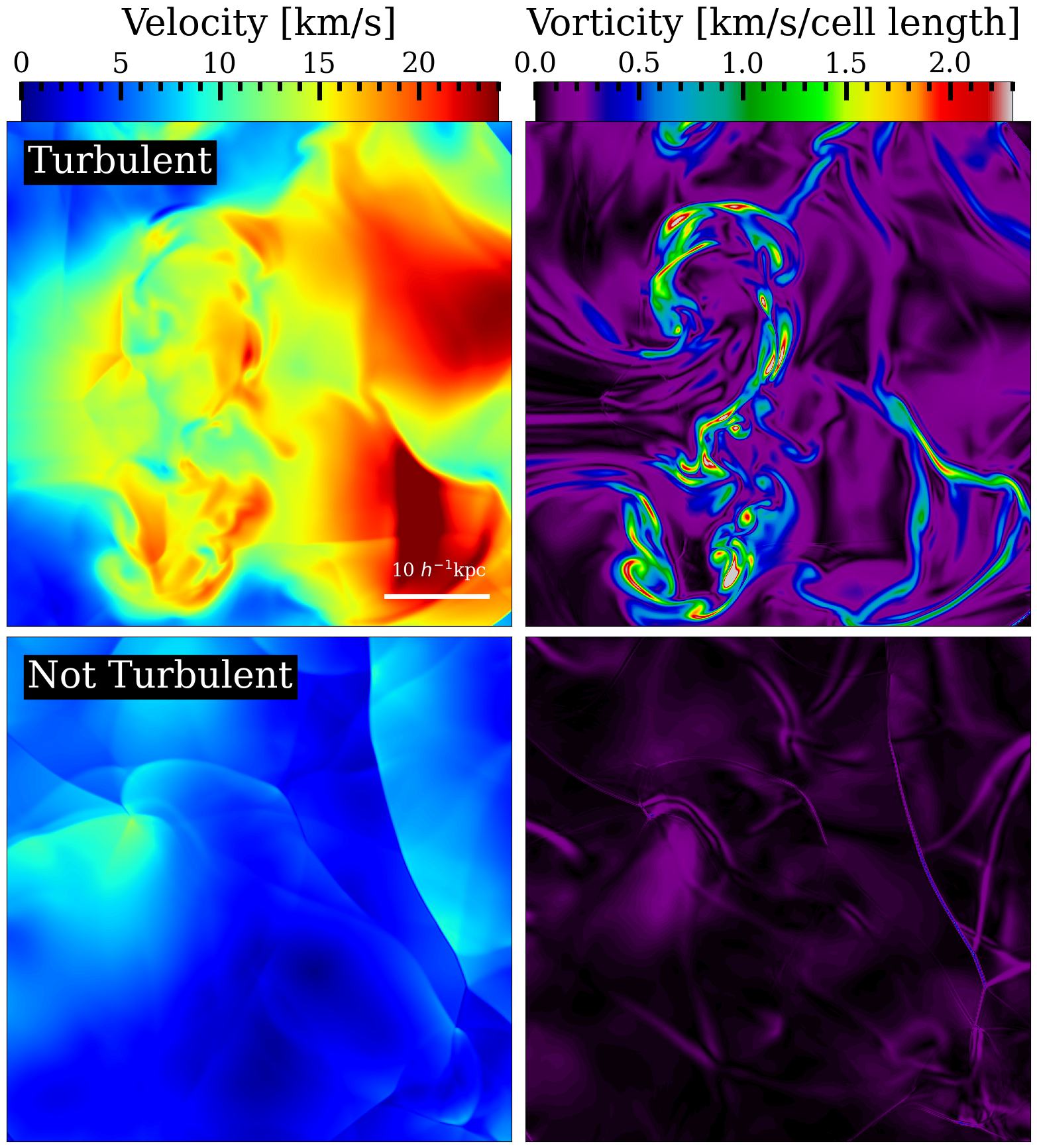}
    \caption{Visualization of the association of turbulence with vorticity.  We show velocity (left) and vorticity (right) in $50$ $h^{-1}$kpc zoom-ins from our $T_{\rm min} = 100$ K simulation with $\Delta x = 122$ $h^{-1}$pc. The top row shows a turbulent region, and the bottom row a non-turbulent one.  }
    \label{fig:vorticity}
\end{figure*}

It is well-known that for gas to exhibit turbulent flow, its vorticity (curl of the velocity) must be non-zero. Since the cosmological initial conditions are thought to have zero vorticity, the only way to source vorticity is if pressure and density gradients are not aligned, which occurs whenever pressure is not a unique function of density. This happens in our simulations after re-ionization as a direct results of the photo-heating and photoevaporation by cosmological ionization fronts~\citep{Hirata2018,Cain2024a}.

We show in Figure~\ref{fig:vorticity} that regions with high vorticity are, indeed, the sights of turbulence in our simulations.  The top row shows a $50$ $h^{-1}$kpc zoom-in on a turbulent region in our $T_{\rm min} = 100$ K simulation with $\Delta x = 122$ $h^{-1}$pc.  The left panel shows the magnitude of the gas velocity, and right shows the vorticity.  The bottom row shows the same thing, but zoomed in on a region that is visibly not turbulent.  We see that the turbulent case has high vorticity in regions where the gas flows are noticeably turbulent on the left.  The gas velocities are also $\sim 20$ km/s here, near the sound speed.  The non-turbulent region has somewhat slower velocity and small (but non-zero) vorticity.  This confirms the expectation that turbulent regions should be associated with high-vorticity regions, and suggests that vorticity may be a useful metric for estimating the volume-filling factor of turbulence robustly, which we plan to study more carefully in future work.  

\subsubsection{Sensitivity to modeling assumptions}
\label{subsec:modeling}

Our fiducial setup makes several modeling assumptions that may significantly affect the emergence of turbulence.  As we discussed in \S\ref{subsec:Jeans}, the presence of small-scale structure near the driving scale is expected to be a condition for turbulence to emerge.  Perhaps most notably, the (highly uncertain) temperature of the pre-ionized gas sets the Jeans scale prior to reionization, and could plausibly prevent turbulence from forming if it were high enough.  Our simulations assume adiabatic cooling prior to reionization without any heating from X-rays, which maximizes the amount of small-scale structure able to form.  This scenario is already ruled out at high confidence by 21 cm observations~\citep{Hera2022}, which require $T_{\rm gas} \gtrsim 10$K at $z \approx 8$.  There are also several other modeling uncertainties that affect the formation of small-scale structure before reionization, which we study here.  

\begin{figure}
    \centering
    \includegraphics[scale=0.275]{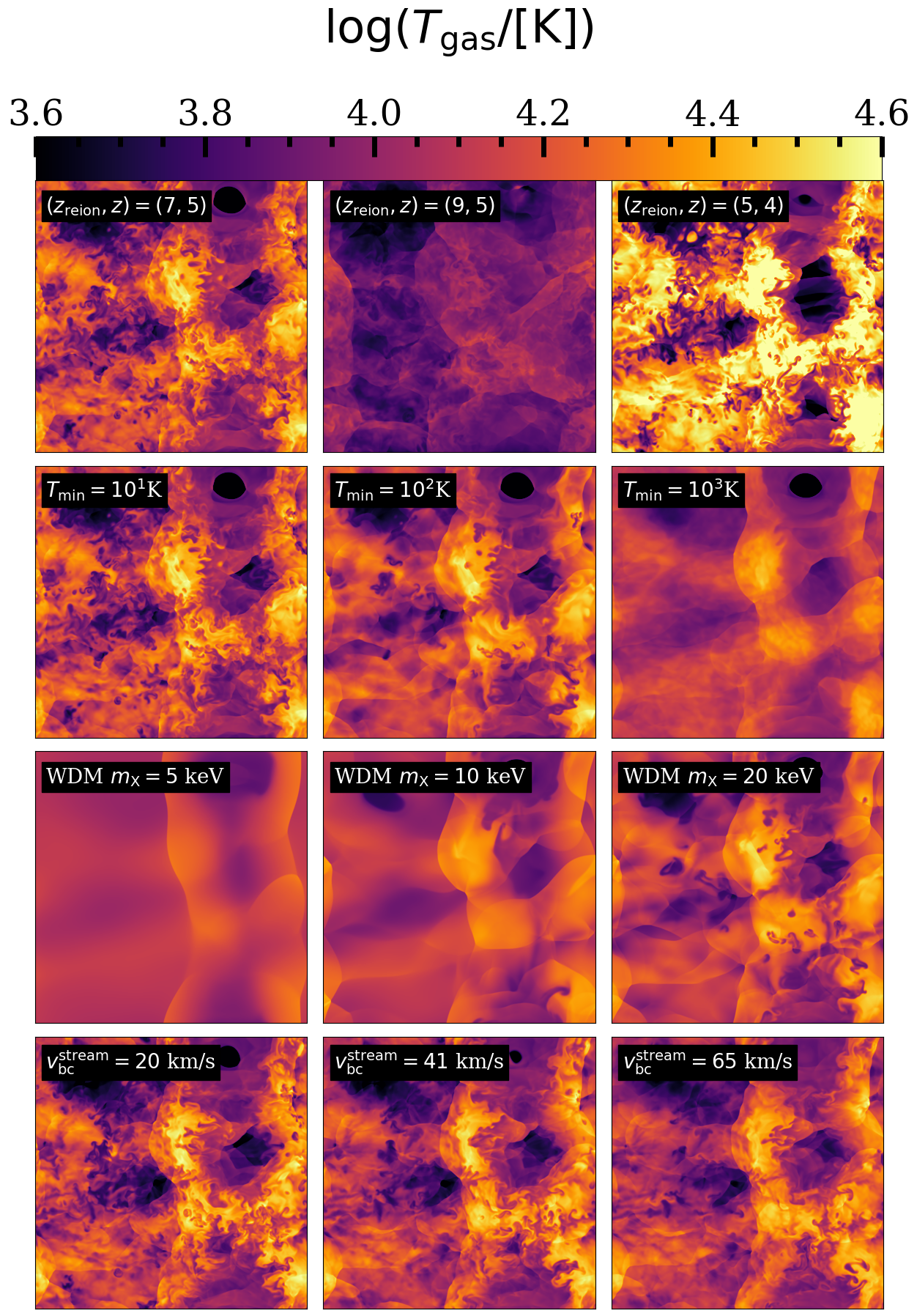}
    \caption{Effect of different box scale physics on the emergence of IGM turbulence - all panels are shown at $z = 5$, unless otherwise indicated.  {\bf Top row}: effect of the reionization redshift.  Higher (lower) $z_{\rm re}$ results in less (more) turbulence, since more small-scale structure has formed in the neutral IGM at lower $z$.  {\bf 2nd row}: effect of X-ray pre-heating.  From left to right, we impose pre-ionization minimum temperatures of $T_{\rm min} = 10$, $100$, and $1000$ K at $z < 15$ to crudely mimic the effects of pre-heating. 
    We see no difference with the unheated case for $T_{\rm min} = 10$ K, but turbulence is greatly reduced for $T_{\rm min} = 100$ K and eliminated entirely for $T_{\rm min} = 1000$ K.  {\bf 3rd row}: effect of a small-scale cutoff in the dark matter power spectrum, parameterized by a WDM particle mass $m_X$ (see text).  Turbulent features are missing entirely even for $m_{\rm X} = 20$ keV.  {\bf Bottom row}: effect of the baryon-dark matter streaming velocity (see text).  For $v_{\rm bc}^{\rm stream} = 20$ km/s, there is little difference with the $v_{\rm bc}^{\rm stream} = 0$ case, but higher $v_{\rm bc}^{\rm stream}$ leads to significantly lower levels of turbulence.  }
    \label{fig:modeling_effects_visualization}
\end{figure}

Figure~\ref{fig:modeling_effects_visualization} shows gas temperature slices at our fiducial resolution, each row varying a different simulation parameter (at $z = 5$, unless otherwise indicated).  In the top row, we vary the redshift of reionization, $z_{\rm re}$, and display the temperature map $\approx 300$ Myr after reionization (motivated by our picture that the turbulence is driven by how far sound waves can travel after ionization).  The first panel shows our fiducial $z_{\rm re} = 7$, and the middle and right panels show $z_{\rm re} = 9$ and $5$, respectively.  We see that turbulence is noticeably less (more) prominent in the $z_{\rm re} = 9$ ($5$) case.  This is likely because denser, colder structures are able to form when the IGM re-ionizes at lower redshift, resulting in a more violent relaxation that drives turbulence more effectively.  The gas at lower redshifts is also less dense and inverse Compton cools less quickly, such that the IGM takes more time to reach thermal and pressure equilibrium.  Recent work supporting a late end to reionization~\citep{Kulkarni2019,Nasir2020,Bosman2021,Zhu2024} suggests that as much as half of the IGM may have been neutral at $z = 7$ and up to $20\%$ at $z = 6$.  

In the second row of Figure~\ref{fig:modeling_effects_visualization}, we show the effect of X-ray pre-heating of the pre-ionized IGM.  We crudely model this by imposing a minimum temperature of $T_{\rm min}$ at $z \leq 15$ prior to reionization, following~\citet{DAloisio2020}.  From left to right, we show $T_{\rm min} = 10$, $100$, and $1000$ K for our fiducial $z_{\rm re} = 7$ at $z = 5$.  The left panel in the second row is almost indistinguishable from the fiducial case above it, suggesting that heating the IGM to $10$ K has almost no effect on turbulence as an unheated IGM.  This is comparable to the current lower limits on the pre-heated IGM temperature from HERA at $z \approx 8$~\citep{Hera2022}.  However, for $T_{\rm min} = 100$ K, turbulent features are considerably suppressed, although not entirely erased.  This is consistent with our argument in \S\ref{subsec:Jeans} that pre-ionization gas temperatures much above $\approx 100$ K would prevent the seeding of turbulence. (One caveat is that our crude implementation of pre-heating -- a temperature floor at $z < 15$ - may artificially increase its importance.  Pre-heating likely starts later than this, such that for a given $T_{\rm min}$ at $z_{\rm re}$, more small-scale structure forms and survives to $z_{\rm re}$ than we find in our simulations.)  For $T_{\rm min} = 1000$ K, there is no sign of any turbulence.   

In the third row, we study the effect that our assumptions about dark matter (DM) have on our results.  In particular, gas clumping on small scales that is needed to seed turbulence will be absent if dark matter does not also clump down to those scales.  As a proxy for such scenario, we initialize several simulations with the warm dark matter (WDM) model of~\citep[][see also~\citealt{Cain2022a}]{Viel2005}, which introduces a cutoff in the initial DM power spectrum that depends on the assumed WDM thermal relic particle mass, $m_{\rm X}$.  We show results for $m_{\rm X} = 5$, $10$, and $20$ keV, of which the last two are well above the most recent upper limits on $m_{\rm X}$ from the Ly$\alpha$ forest~\citep{Villasenor2023,Irsic2024} and gravitational lensing~\citep{Gilman2019}.  The free-streaming scale, below which the matter power spectrum is cut off, is $k_{\rm FS} \approx 37$, $74$, and $147$ $h$Mpc$^{-1}$, respectively (Eq. 3 of~\citet{Viel2005}).  We see no clear signs of turbulence in the $5$ or $10$ keV models.  The $20$ keV case displays a small amount of turbulence close to the largest filaments, and appears very similar to the $T_{\min} = 10^2$ K model.  This signals that IGM turbulence is very sensitive to the DM power spectrum down to scales currently not probed directly by observations.  Conclusive evidence for such turbulence may place very sensitive limits on alternative DM cosmologies.  

The last row shows the effect of the velocity offset between baryons and dark matter (streaming velocity, $v_{\rm bc}^{\rm stream}$) induced at recombination~\citep{Tseliakhovich2010}.  This effect suppresses the formation of structure on small scales by increasing the minimum DM halo mass required to accrete gas from the IGM.  It has been shown to have a potentially significant effect on Pop III star formation and the 21 cm  signal~\citep{Munoz2019} and may affect the IGM even during reionization~\citep{Cain2020,Park2021}.  We implement $v_{\rm bc}^{\rm stream}$ following the procedure used in~\citep{Cain2020}.  We show results for $v_{\rm bc}^{\rm stream} = 20$ km/s, $41$ km/s, and $65$ km/s at $z = 1080$, the same values used in~\citet{Cain2020}, which are $0.66$, $1.3$, and $2.2\times$ the RMS value of $\sigma_{\rm bc} = 30$ km/s.  Compared to the upper left-most panel, the case with $v_{\rm bc}^{\rm stream} = 20$ km/s has only slightly reduced turbulence.  However, at higher $v_{\rm bc}^{\rm stream}$, turbulent features become significantly less prominent, although they never disappear entirely.  This raises the intriguing possibility that the strength of IGM magnetic fields may be a biased tracer of the large-scale $v_{\rm bc}^{\rm stream}$ field, which exhibits strong Baryon Acoustic Oscillation (BAO) features that are offset relative to the standard BAO used as a standard ruler~\citep{2016PhRvL.116l1303B}.  If these magnetic fields then seed galactic magnetic fields, which then perhaps affect star formation, this could even result in a bias in the BAO peak position in galaxy clustering measurements.

\begin{figure*}
    \centering
    \includegraphics[scale=0.3]{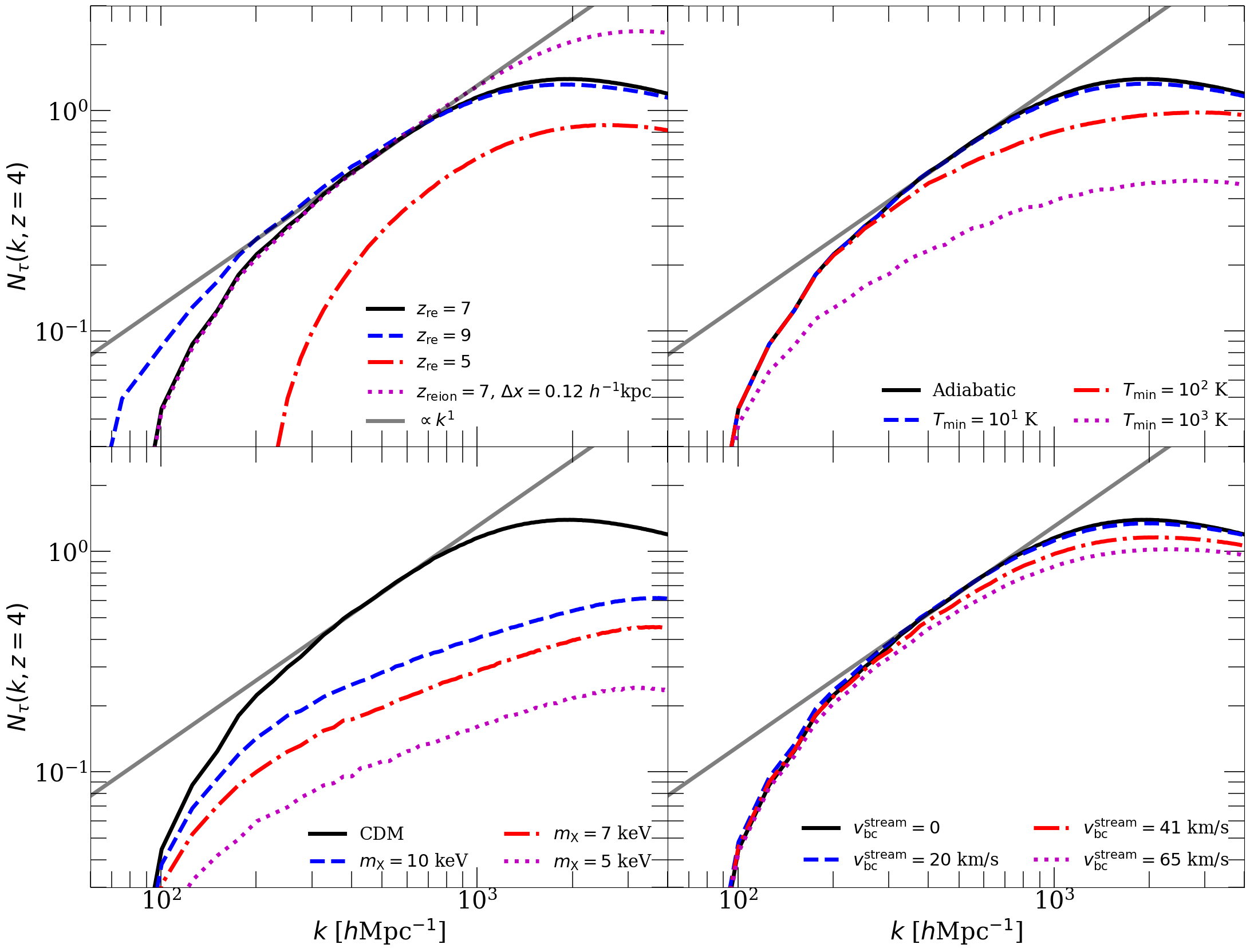}
    \caption{Estimate of the number of eddy turnover timescales elapsed by $z = 4$ as a function of scale.  {\bf Top Left:} comparison of different $z_{\rm re}$ and resolutions.  The integrated $N_{\tau}$ is similar in the $z_{\rm re} = 7$ and $9$ cases, while in the $z_{\rm re} = 5$ case relaxation is still likely ongoing and driving turbulence.  We find $N_{\tau} \propto k$ at scales where turbulence is observed in Figure~\ref{fig:ke_scaling}, and the high-resolution run traces this scaling up to higher $k$ than the fiducial case.  {\bf Top Right:} dependence on $T_{\min}$.  There is little difference with the adiabatic case for $T_{\min} = 10^1$K, but $N_{\tau}$ decreases and deviates from the expected scaling at higher $T_{\min}$.  {\bf Bottom Left:} dependence on $m_{\rm X}$.  All the WDM cases fall well below the CDM case and do not have the expected scaling for turbulence, consistent with Figure~\ref{fig:modeling_effects_visualization}.  {\bf Bottom Right:} dependence on $v_{\rm bc}^{\rm stream}$.  We see much milder dependence here than in the other panels, with $N_{\tau}(k)$ flattening slightly at high $k$ for large $v_{\rm bc}^{\rm stream}$.  }
    \label{fig:timescales_models}
\end{figure*}

Next, we make these comparisons quantitative by estimating the number of eddy turnover times (Eq.~\ref{eq:tau_eddy}) elapsed in each simulation by $z = 4$.  This is given by
\begin{equation}
    N_{\rm tau}(k,z=4) = \int_{z_{\rm re}}^{4} dz \frac{dt}{dz} \frac{1}{\tau_{\rm eddy}(k)}
\end{equation}
We show this quantity in Figure~\ref{fig:timescales_models} for the same simulations shown in Figure~\ref{fig:modeling_effects_visualization}.  The top left panel compares simulations with $z_{\rm re} = 5$, $7$, and $9$, and our high-resolution ($\Delta x = 0.12$ $h^{-1}$) run with $z_{\rm re} = 7$.  We see that in the range of scales we previously identified as turbulent for our fiducial simulation ($300 \lesssim k/[h{\rm Mpc}^{-1}] \lesssim 1000$), we find that $N_{\tau}$ scales approximately like $\propto k^1$, steeper than the $k^{2/3}$ scaling expected for the $E(k) \propto k^{-5/3}$ Kolmogorov scaling (see Eq.~\ref{eq:tau_eddy}).  The reason for this is likely that the sound crossing time for gas at scale $k$, $2\pi/k/c_s$, decreases towards higher $k$, allowing more time for turbulence to form at smaller scales just after reionization, when the driving force is strongest.  This steepens the $k$ dependence of $N_{\tau}$ because smaller scales become turbulent earlier than larger ones.  

We see that our $z_{\rm re} = 9$ and $7$ runs yield very similar values of $N_{\tau}$ despite reionizing at different times.  This is probably because the turbulence only lasts as long as the driving force is in effect, and this timescale does not vary strongly with redshift.  The similarity of the $z_{\rm re} = 9$ and $7$ cases suggests that $N_{\tau}$ may be roughly independent of $z_{\rm re}$ when integrated to low enough redshift.  In the $z_{\rm re} = 5$ case, the driving force probably has not dissipated yet, such that we would find a larger $N_{\tau}$ if we could integrate to lower redshift.  For the $z_{\rm re} = 7$ and $9$ cases, we find that $N_{\tau} \approx 1$ at our fiducial resolution and $k \approx 800$ $h$Mpc$^{-1}$, the highest $k$ for which $N_{\tau}$ traces the gray line.  The comparison to our high-resolution run shows that $N_{\tau}$ continues to trace this line up to higher scales when resolution increases, which is encouraging for the prospect of B-field amplification, as we argued in \S\ref{sec:Bfieldgrowth}.  

The top right panel shows the effect of varying $T_{\min}$.  As expected from Figure~\ref{fig:modeling_effects_visualization}, we find little difference between the adiabatic run and $T_{\min} = 10$K.  The $T_{\min} = 10^2$K case deviates from the $\propto k$ appreciably, and the $T_{\min} = 10^3$K case falls well below $N_{\tau} = 1$ at all scales.  This confirms our earlier finding that turbulence is suppressed at $T_{\min} = 10^2$K and disappears for pre-heating rates much higher than this.  We see something similar to the $T_{\min} = 10^3$K case in the lower left panel, which shows results for different WDM particle masses.  None of the curves reach $N_{\tau} = 1$ or match the scaling seen in the CDM case.  The lower right panel shows that $v_{\rm bc}^{\rm stream}$ has a relatively small effect on the inferred $N_{\rm tau}$, but does modestly change its scaling with $k$.

\subsection{Sensitivity to the self-shielding treatment}
\label{app:self_shielding}

\begin{figure*}
    \includegraphics[scale=0.28]{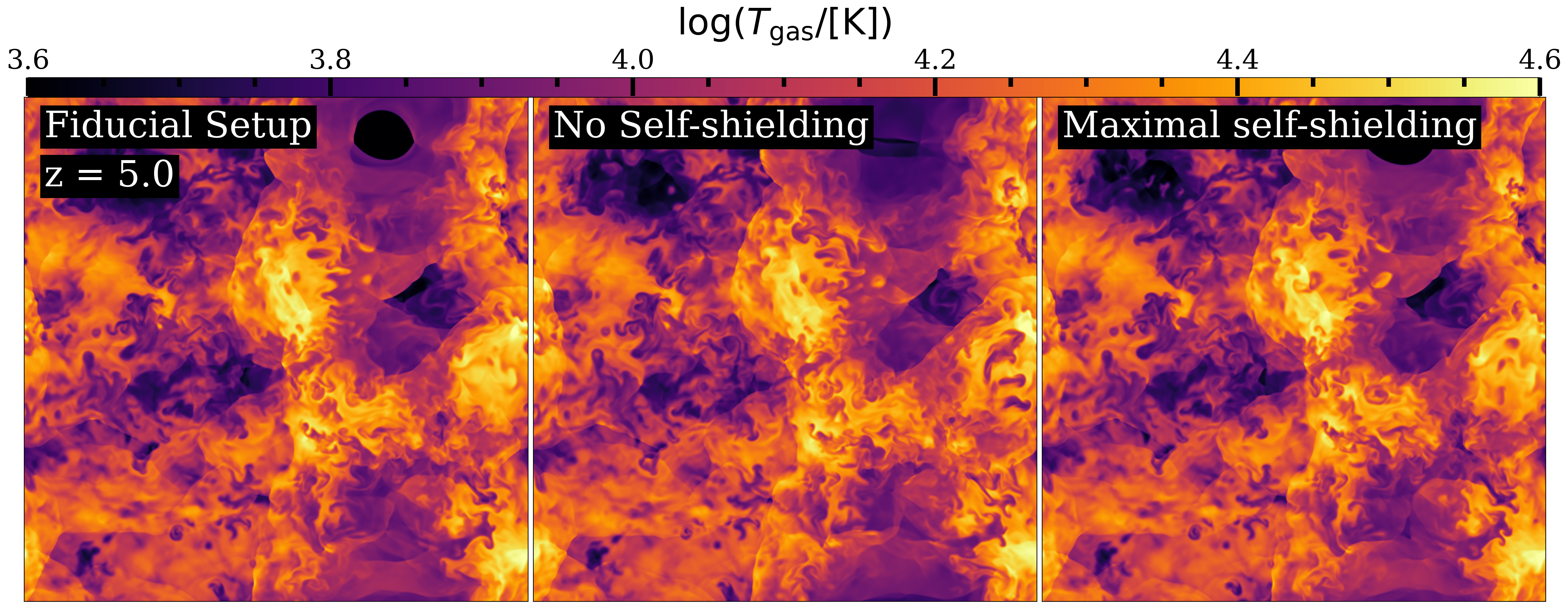}
    \caption{Effect of different self-shielding implementations on IGM turbulence. 
 The left panel shows a temperature slice at $z = 5$ using our fiducial setup, and the middle panel shows the same for a flash-ionized run at the same resolution.  The run in the right panel did not include our density-freezing procedure, and the I-front came from only one direction and traversed the entire box -- thus maximizing self-shielding.  We find no appreciable difference between these cases, showing that our self-shielding implementation does not affect our main results.  }
    \label{fig:self_shielding}
\end{figure*}

We test here whether our findings are sensitive to the implementation of self-shielding in our simulations.  Throughout this work, we use the same setup for sending I-fronts across RT domains described in~\citet{Cain2024a}.  At $z_{\rm re}$, we begin tracing rays from the boundaries of RT domains, and ``freeze'' both cosmic time and hydrodynamical evolution until the I-fronts have crossed the RT domains.  This ensures that the gas throughout the box responds coherently to the effects of reionization, and that the volume is fully reionized at exactly $z_{\rm re}$.  During this time, the reduced speed of light is set to $5\times$ the speed of the I-front crossing the RT domain, or its full value, whichever is smaller, to ensure the correct I-front speed and post-I-front gas temperature~\citep{DAloisio2019,Zeng2021}.  At $z < z_{\rm re}$, we use a reduced speed of light of $\tilde{c} = 2 \times 10^{-3}$, which we find does not affect the self-shielding properties of the gas significantly\footnote{Because of the tiny light-crossing time at the cell size, these simulations become computationally intractable if the speed of light is not reduced significantly.  }.  We show here that this procedure does not affect our qualitative results, but does simplify their interpretation.  

We have run two additional simulations with $N = 1024^3$.  The first flash-ionizes the box at $z = z_{\rm re}$ (just like our $N = 2048^3$) run, and thus has no self-shielding.  The second sends a single plane-parallel I-front across the entire box at $z = z_{\rm re}$ from just one direction, and does not freeze the density field during I-front passage.  Respectively, these two approaches minimize and maximize the effect of self-shielding on IGM gas dynamics.  

We show slices through the temperature field at $z = 5$ for each of these cases in Figure~\ref{fig:self_shielding}.  The turbulent eddies are clearly visible in all three cases, and there is very little discernible difference between them.  This demonstrates that our main conclusions are insensitive to the details of how self-shielding is handled.  The only significant difference in Figure~\ref{fig:self_shielding} is that the middle panel does not have the conspicuous ``hole'' in the upper right corner of the map.  This region was occupied by a dense clump before reionization, which was rarefied so much that the leftover under-density was cooled all the way to a few hundred K by $z = 5$.  The destruction of this clump seems to have occurred with different timing in the flash-ionized run, causing the local temperature and density gradients to be less steep at lower redshifts.   \\

\subsection{Test with Nyx}
\label{app:Nyx}

\begin{figure*}
    \centering
    \includegraphics[scale=0.28]{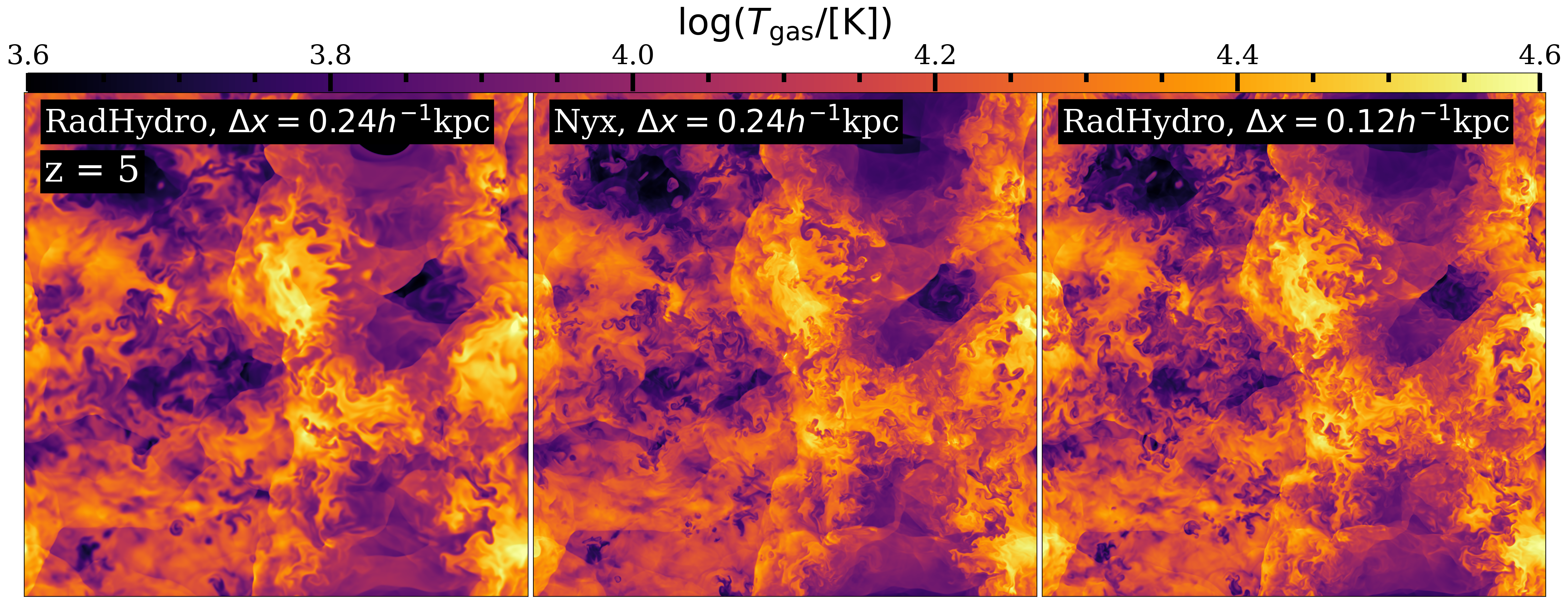}
    \caption{Comparison of IGM turbulence in RadHydro and Nyx.  We show slices through the temperature map at $z = 6$, $5$, and $4$ (left to right) for our fiducial RadHydro run, a Nyx run with matching resolution, and our highest resolution RadHydro run (top to bottom).  We find qualitatively similar turbulence in Nyx to that observed in RadHydro.  Indeed, Nyx shows more turbulence on smaller scales at a fixed resolution than RadHydro, suggesting its algorithm is less diffusive.  }
    \label{fig:RH_Nyx_comparison}
\end{figure*}

Similar to Appendix A of~\citet{Cain2024a}, we have run a test comparing our RadHydro results to a similar simulation setup run with the Nyx code~\citep{Almgren2013}.  In Figure~\ref{fig:RH_Nyx_comparison}, we show slices through the temperature fields from our Nyx run and two RadHydro runs at $z = 5$.  The left panel shows our fiducial RadHydro run, the middle a Nyx run with matching resolution ($0.24$ $h^{-1}$kpc), and the right our $0.12$ $h^{-1}$kpc resolution RadHydro run.  The features of the maps on large scales are very similar between RadHydro and Nyx, and the Nyx run clearly displays the turbulence, as seen in RadHydro, evidencing the numerical robustness of our findings.  

Strikingly, the Nyx run displays more turbulent structure than the fiducial RadHydro run and with smaller-scale structure.  The features in the Nyx run are actually more similar to the RadHydro run at higher resolution in the right-most panel.  This could be because of algorithmic differences between RadHydro and Nyx that affect diffusivity on small scales~\footnote{There are several differences between the numerical methods used in RadHydro and Nyx that could contribute to this result.  These include different hydrodynamical solvers (relaxing TVD~\citep{Trac2003} vs. iterative Riemann), different treatments of gravity (FFTs vs. Gauss-Seidel relaxation), and differing treatments of the equation of state prior to reionization.  A suite of tests involving these and other codes would be needed to pin down the source of the differences.  }.  A higher numerical Reynolds number (see Eq.~\ref{eq:Renum}) at fixed cell size would reduce the minimum scale at which turbulence can appear (Eq.~\ref{eq:resolution}), which is what we see here.  This suggests that forthcoming studies of reionization-driven turbulence should take into consideration the hydrodynamical solver employed. 

\end{document}